\documentclass[journal]{IEEEtran}
\usepackage{subfigure}
\usepackage[pdftex]{graphicx}
\usepackage{epstopdf}
\usepackage{xcolor}
\usepackage{amsmath}
\usepackage{mathtools}
\usepackage{amssymb}
\usepackage{verbatim} 
\usepackage{diagbox}

\usepackage{ifpdf}

\ifCLASSINFOpdf
\else
\fi
\begin{document}
%
\title{Predicting the Quality of Compressed Videos with Pre-Existing Distortions}
%
%
%

 \author{Xiangxu Yu, Neil Birkbeck, Yilin Wang, Christos G. Bampis, Balu Adsumilli and Alan C. Bovik
 \thanks{X. Yu and A. C. Bovik are with the Department of Electrical and Computer Engineering, University of Texas at Austin, Austin, USA (e-mail: yuxiangxu@utexas.edu;  bovik@ece.utexas.edu). Neil Birkbeck, Yilin Wang and Balu Adsumilli are with YouTube Media Algorithms Team, Google Inc (e-mail: birkbeck@google.com; yilin@google.com; badsumilli@google.com). C. G. Bampis is with Netflix Inc (e-mail: cbampis@gmail.com).}
 }

\markboth{----------------------------------------------}%
{--------------------------------}
%



\maketitle

\begin{abstract}
Over the past decade, the online video industry has greatly expanded the volume of visual data that is streamed and shared over the Internet. 
Moreover, because of the increasing ease of video capture, many millions of consumers create and upload large volumes of User-Generated-Content (UGC) videos. 
Unlike streaming television or cinematic content produced by professional videographers and cinemagraphers, UGC videos are most commonly captured by naive users having limited skills and imperfect technique, and often are afflicted by highly diverse and mixed in-capture distortions. 
These UGC videos are then often uploaded for sharing onto cloud servers, where they further compressed for storage and transmission. 
Our paper tackles the highly practical problem of predicting the quality of compressed videos (perhaps during the process of compression, to help guide it), with only (possibly severely) distorted UGC videos as references. 
To address this problem, we have developed a novel Video Quality Assessment (VQA) framework that we call 1stepVQA (to distinguish it from two-step methods that we discuss). 
1stepVQA overcomes limitations of Full-Reference, Reduced-Reference and No-Reference VQA models by exploiting the statistical regularities of both natural videos and distorted videos. 
We show that 1stepVQA is able to more accurately predict the quality of compressed videos, given imperfect reference videos. 
We also describe a new dedicated video database which includes (typically distorted) UGC reference videos, and a large number of compressed versions of them. 
We show that the 1stepVQA model outperforms other VQA models in this scenario. 
We are providing the dedicated new database free of charge at https://live.ece.utexas.edu/research/onestep/index.html
\end{abstract}


%
\IEEEpeerreviewmaketitle

\section{Introduction}
\label{introduction}

\IEEEPARstart{I}{n} recent years, digital images and videos have become remarkably ubiquitous and now constitute the majority of Internet traffic. 
According to \cite{ciscociscovisualnetworkingindex}, video streaming continues to occupy a growing share of Internet bandwidth, and by 2022, it is expected that 82\% of global ``moving bits'' will be picture and video content. 
A substantial portion of this data is generated by streaming providers like Netflix, Hulu, and Amazon Prime Video. 
The content they provide, with some exceptions, has been created by expert photographers using professional capture devices. 
These pristine videos are generally (but not always) of high quality, and can usually be used as references in subsequent Video Quality Assessment (VQA)/compression processes.  
However, another category of videos are also uploaded and downloaded in gigantic volumes by casual users, called User-Generated-Content (UGC) videos. 
These imperfect videos are very often uploaded onto social platforms like YouTube, Snapchat, Facebook, Instagram, and TikTok. 
UGC content is commonly captured by inexpert using having uncertain technique and often unsteady hands, resulting in numerous, often commingled impairments of perceived quality. 
These UGC videos have often undergone a series of processing steps, such as editing, aesthetic modification, and compression, before the user uploads them to an online server, where they inevitably undergo \textit{another} round of compression. 
The highly diverse mixtures and severities of distortion that these UGC videos contain are very difficult to model. 
Since high quality pristine reference videos cannot be counted on to guide compression, it is highly desirable to find ways of assisting compression decisions by accounting for the qualities of the source videos. 
This challenging problem presents unique difficulties in perceptual distortion modeling. 

We remind the reader that VQA models can be conveniently placed into two categories. 
Those that require a reference signal, which includes Full-Reference (FR) and Reduced-Reference (RR) models, and those that do not: No-Reference (NR) models. 
Current mainstream reference VQA models include the FR VMAF \cite{li2016toward}, VQM \cite{pinson2004new}, SSIM \cite{wang2004image} and VIF \cite{sheikh2006image}, and the RR model ST-RRED \cite{soundararajan2012video}. 
The second main category is that of NR models, which aim to accurately predict picture quality without the aid of any reference videos. 
In applications involving UGC videos, reference models (FR or RR) are problematic, since comparing a possibly distorted test video against a reference that is also distorted must lead to errors, and possibly severe losses of quality prediction accuracy. 
While NR models are intended to be able to predict the quality of distorted-then-compressed videos, NR VQA remains a challenging research topic and the most competitive models still have difficulty in predicting the quality of complex distorted UGC videos \cite{sinno2018large}. 

Here we seek to advance progress on solving the problem of predicting the quality of distorted-then-compressed videos. 
For example, in a cloud server, many distorted and already-compressed videos will be uploaded by users, only to be further compressed without the benefit of a guiding pristine reference. 
Instead, the source video may already contain any of many kinds of possibly mixed distortions, including almost-inevitable prior compression. 
Thus, standard reference VQA models cannot be expected to produce optimal quality predictions. 
To address this issue, we propose a method called 1stepVQA, which learns to predict compressed UGC video quality, including relationships between native distorted UGC video quality and that of further video compression. 
1stepVQA accomplishes this by monitoring losses of expected statistical regularity in videos being analyzed. 
Since distortions can cause the statistical properties of videos to deviate from well-modeled natural regularities, it is possible to learn relationships between distortions and ``natural video statistics'' and use them to predict perceptual quality. 
In the approach taken here, we seek to avoid the drawbacks of reference VQA models by also making measurements of quality-aware features on the typically flawed reference videos. 

The essential tools needed to solve VQA problems are databases of appropriate diverse video contents and distortions, labeled with adequate amounts of subjective data. 
There are a number of legacy VQA databases, including LIVE VQA \cite{seshadrinathan2010study}, LIVE Mobile \cite{moorthy2012video}, CDVL \cite{pinson2013consumer} and MCL-V \cite{lin2015mcl}, MCL-JCV \cite{wang2016mcl} and VideoSet \cite{wang2017videoset}, each containing about 10-20 pristine video contents, and many distorted versions of them. 
The distortions in all of these databases were synthetically generated in isolation.  
While these databases have successfully driven the development of early reference VQA models, they are of limited value with respect to the UGC video quality assessment problem. 
More recently, a few novel databases have been developed, including LIVE VQC \cite{sinno2018large}, KoNViD-1k \cite{hosu2017konstanz} and YouTube UGC \cite{wang2019youtube}, which contain many UGC videos of very diverse contents and distortions. 
These databases were originally designed for the development of NR VQA models, but we will also find them useful here. 
The source videos were collected from diverse users and online repositories. 
These databases are not of direct value, and so it was necessary for us to generate a new and dedicated subject VQA database, containing UGC videos of highly diverse real-world content and distortions as source ``reference,'' as well as additionally compressed versions of them, to help us develop and evaluate our model. 

The rest of the paper is organized as follows: we introduce related work in Section \ref{related_work}. 
We present our newly built database in Section \ref{database}. 
We describe our proposed model in Section \ref{1stepVQA_model}. 
We examine our model performance in Section \ref{performance_and_analysis}. 
Finally we summarize the paper in Section \ref{conclusion}.

\section{Related Work}
\label{related_work}

Early work on VQA models largely focused on reference-based models, beginning with the simple PSNR and advancing to more sophisticated FR frame-based models that better account for visual perception, including SSIM, MPQM \cite{van1996perceptual}, MS-SSIM \cite{wang2003multiscale}, VIF, VSNR \cite{chandler2007vsnr}, FSIM \cite{zhang2011fsim}, IW-SSIM \cite{wang2010information}, and DOG-SSIM \cite{pei2015image}, and spatial-temporal models like VMAF, VQM, MOVIE \cite{seshadrinathan2009motion}, ST-MAD \cite{vu2011spatiotemporal}, TetraVQM \cite{barkowsky2009temporal}, and an optical flow-based method \cite{manasa2016optical}. 
There are also successful RR models, such as ST-RRED and SpEED-QA \cite{bampis2017speed}. 

The development of NR models which do not require any reference videos has been a hot topic in recent years, driven by the need to evaluate UGC videos and older, lower-quality content. 
Early models were often distortion-specific, targeting only one or more specific distortions. 
Later, more general models have involved training quality prediction models to learn mappings from features to subjective judgments, e.g., BRISQUE \cite{mittal2012no}, V-BLIINDS \cite{saad2014blind} and BIQME \cite{gu2017learning}, which make use of natural scene statistics (NSS) models; BPRI \cite{min2016blind,min2018blind}, which utilizes a ‘pseudo reference image’, TLVQM \cite{korhonen2019two}, which deploys a simplified motion estimator, and more recent deep learning models, like NIMA \cite{talebi2018nima}, PaQ-2-PiQ \cite{ying2019patches}, PQR \cite{zeng2018blind}, DLIQA \cite{hou2015blind}, and more \cite{kim2017deep}. 
There are also completely blind (unsupervised) models, like NIQE \cite{mittal2013making} and IL-NIQE \cite{zhang2015feature}. 

The approach that we take to the problem of assessing the quality of distorted videos undergoing compression relies on spatial-temporal NSS models. 
NSS models are the foundation of many FR, RR and NR models. 
In particular, ``distorted NSS'' models seek to quantify losses of statistical regularity in visual signals. 

In these approaches, bandpass picture or video coefficients are modeled by Gaussian Scale Mixture (GSM) distributions. 
These models typically operate in a bandpass (wavelet, DCT scale-space, etc.) domain. 
In the absence of distortion, perceptually relevant conditioning or divisive normalization nonlinearities tend to Gaussianize and further reduce the redundancy of these coefficients.
Departures from uncorrelated Gaussianity of the coefficients arising from distortion are measured or learned by Image Quality Assessment (IQA)/VQA models, and mapped to perception. 

One of the most commonly used spatial-NSS models, which is closely related to methods used here, is BRISQUE. 
When applied on videos, BRISQUE operates by computing Mean Subtracted Contrast Normalized (MSCN) coefficients on a frame basis. 
The statistics of MSCN of high quality video frames tend to follow a Gaussian distribution, but this characteristic is generally disturbed by the presence of distortions. 
The MSCN of distorted video frames are commonly modeled as following a parametric Generalized Gaussian Distribution (GGD), the parameters of which tend to vary with distortion type and severity. 
During training, MSCN features are computed on a set of distorted videos, from which model parameters are extracted and then fed into a machine learning tool along with human quality labels, to learn mappings from model space to perceived quality. 
The trained model can then be used to blindly predict the quality of videos. 

In \cite{sinno2019spatio}, the authors devised a spatial-temporal VQA model, which utilizes the statistics of directional frame differences. 
These are defined as differences between adjacent frames that have been spatially displaced (by $\pm$ 1 pixel) relative to one another, thereby capturing directional space-time bandpass behavior that exhibits high statistical regularity. 
We build on this model in defining the quality-aware feature set in 1stepVQA. 

The new model we introduce here, called 1stepVQA, utilizes similar underlying models as BRISQUE, but also employs temporal features. 
Moreover, we created a new, special-purpose subjective database to enable a unique training process, whereby two types of human subjective labels are trained on: (i) pre-compressed quality, and (ii) post-compressed quality. 
In this way, the 1stepVQA model is able to learn to predict post-compression video quality, while \textit{also} accounting for pre-compressed quality. 

Previously, we proposed an IQA model \cite{yu2019predicting} to solve the same problem (predicting the quality of still pictures with pre-existing distortions that are then JPEG compressed). 
Unlike 1stepVQA, this prior approach, called 2stepQA, operates in two distinct, sequential, quality assessment stages \cite{bovik2020assessing}. 
The first stage conducts traditional NR IQA (e.g., using NIQE \cite{mittal2013making}) on the source picture, to form a prediction of the pre-existing picture quality. 
In the second stage of 2stepQA, a standard reference IQA model (e.g., SSIM \cite{wang2004image}) is used to predict the post-compressed quality \textit{relative} to the source picture. 
These quality predictions are then combined, using, for example, a simple weighted product of FR/RR and NR stages. 
Instead, 1stepVQA attempts to predict the output compressed quality by simultaneously processing NR quality-aware features on the source and reference VQA features from the input-output video pair. 

The contributions that we make are summarized as follows:
\begin{itemize}
  \item We combine traditional spatial NSS models with novel models of the statistics of displaced temporal frame differences. 
  We are able to show that both modes of statistical data boost performance. 
  We also show that using displaced frame differences rather than non-displaced differences boosts prediction accuracy. 
  \item We created a major new first-of-a-kind subjective VQA database containing 55 unique UGC video contents, along with human quality opinions of them, and many compressed versions of these, on which we also collected a large number of human subjective opinions. 
  \item We used this new psychometric resource to learn a new space-time NSS feature-based VQA model, called 1stepVQA, which utilizes both traditional spatial NSS features, as well as new displaced space-time difference based-NSS features. 
  This new model does not require the overhead of motion estimation, or transformation to another domain. 
  It is relatively fast, uses fewer features than most other VQA models, yet it generates highly competitive results against other top performing models on this important practical problem. 
\end{itemize}
\section{New Database}
\label{database}

In recent years, the number of UGC videos that are uploaded daily has increased to an incredible degree. 
Most of these contributors are amateur videographers having limited skills and uncertain hands, hence, the qualities of these video uploads varies significantly over a very wide range. 
These UGC videos are then uploaded onto Internet servers, and pass through multiple processing stages, before being streamed to potentially millions of clients. 
These uploaded UGC videos will often be corrupted by any of many possible distortions (exposure, shake, multiple types of noise or blur and many more). 
The stages of processing may introduce further defects, including any additional video coding. 
For example, when a video is recorded, it may be compressed within a device before being uploaded. 
After being uploaded to a server, the video may be further compressed for storage and transmission. 

Under this scenario, the uploaded video may already suffer from mixed in-capture distortions, against which there are no reference videos of `pristine' quality, hence FR/RR VQA models may fail if used to predict the quality of the pre-distorted/multiply-compressed video. 
While it would be desirable to directly apply an NR VQA model, to the ultimate compressed output, current NR algorithms are insufficiently general to effectively conduct this very complex task. 
There is also a lack of dedicated databases designed to model this scenario. 
There appears to be a database that may relate to this problem \cite{li2019ugc}, but it is not publicly available. 
Towards filling this gap, we created a new database, called the LIVE Wild Compressed Video Quality Database, which contains hundreds of compressed UGC videos, including the source (reference) videos. 

\subsection{Database Content}
We randomly selected 55 different reference videos (contents) from among the 110 1080p videos contained in the LIVE VQC Database, each of duration 10 seconds. 
These are all UGC videos captured with highly diverse mobile cameras, covering a wide range of contents and qualities. 
Most of these videos are corrupted by diverse authentic, mixed in-capture distortions. 

\begin{figure}
 \centering
 \subfigure[]{
   \label{MOS_distribution_VQC:a}
   \includegraphics[width=0.48\columnwidth]{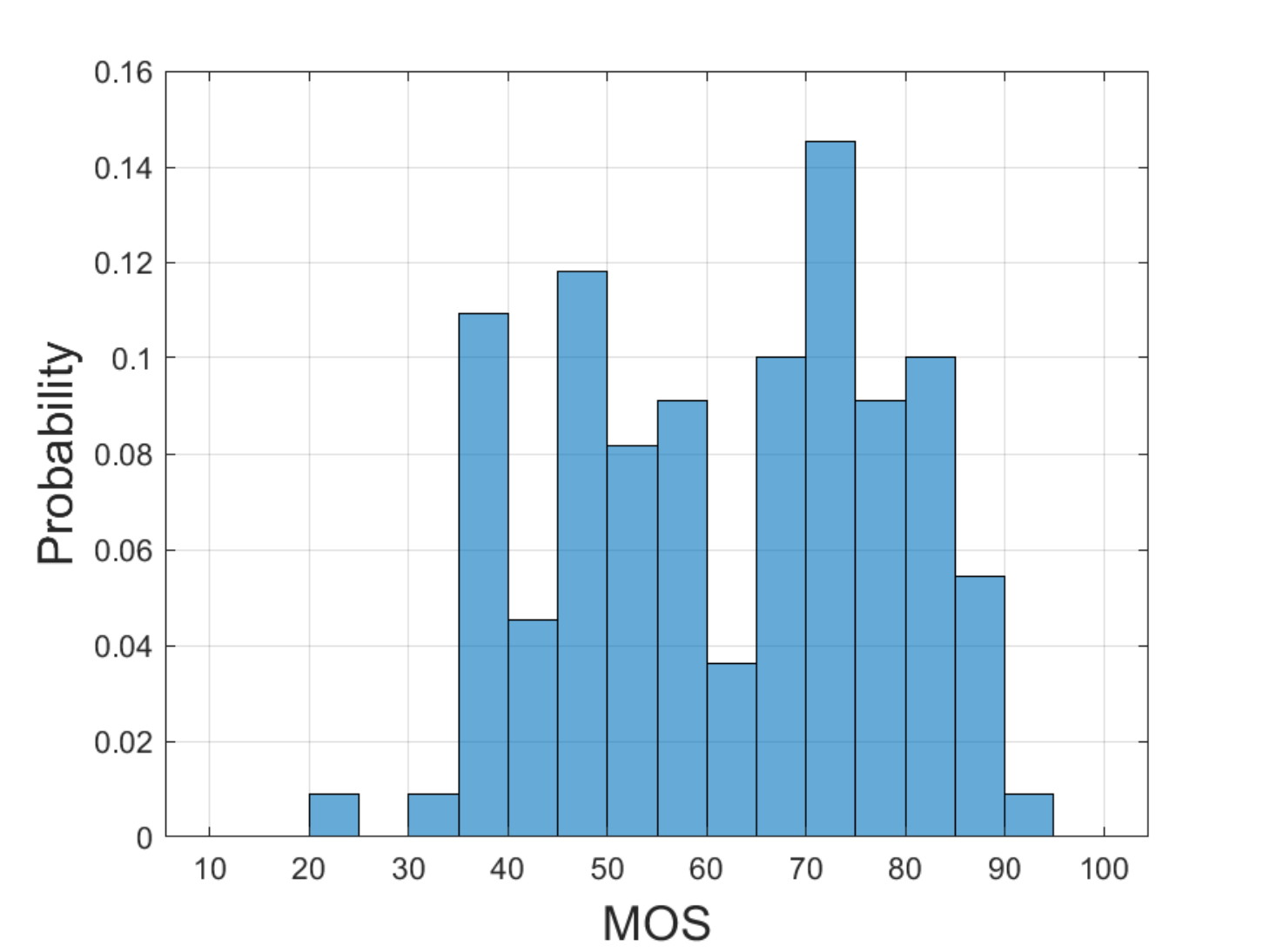}}
 \subfigure[]{
   \label{MOS_distribution_VQC:b} 
   \includegraphics[width=0.48\columnwidth]{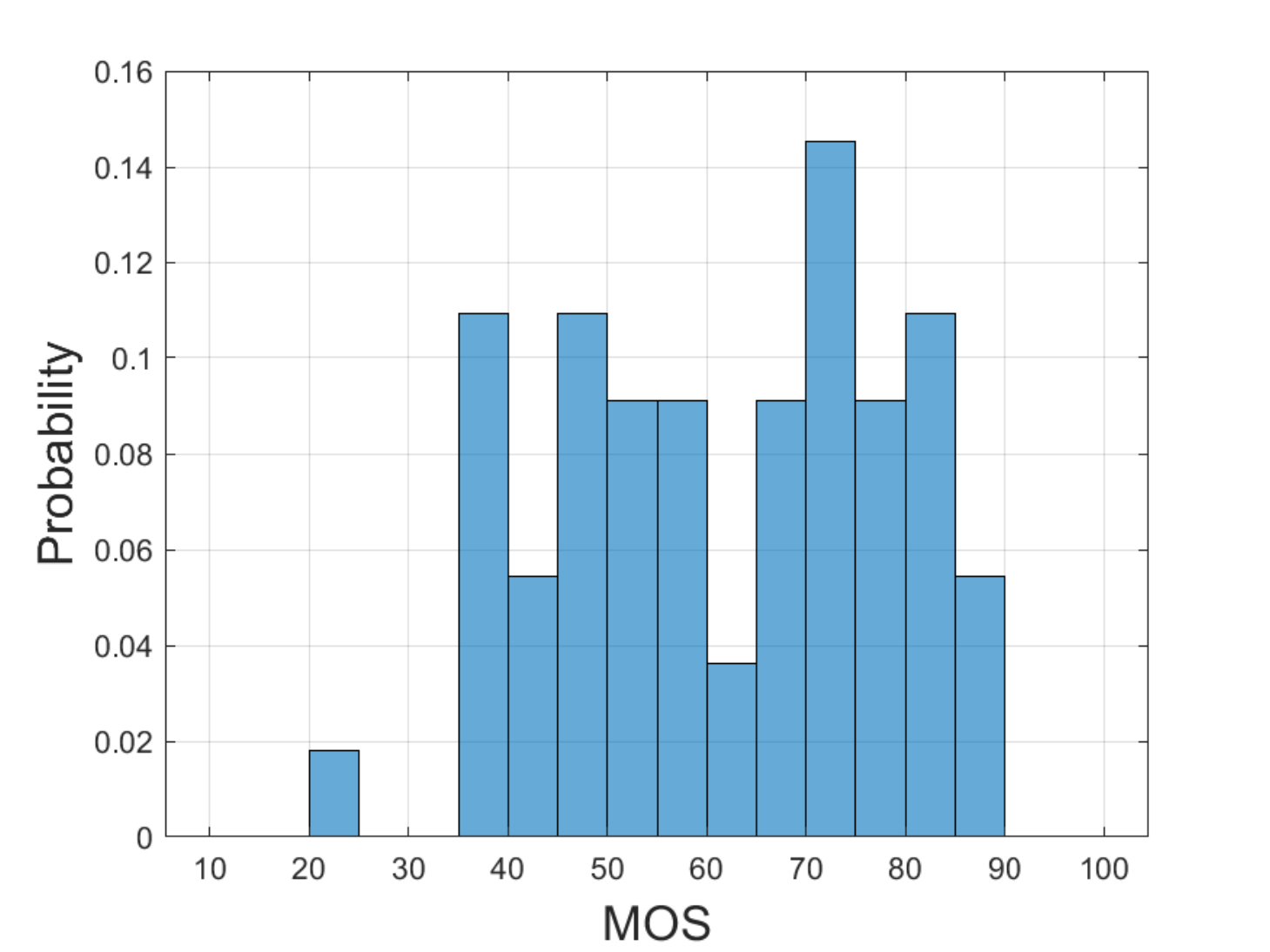}}
 \caption{(a) MOS distribution of the 110 source videos in the LIVE Video Quality Challenge Database. (b) MOS distribution of the 55 selected reference videos in the new LIVE Wild Compressed Video Database.}
 \label{MOS_distribution_VQC} 
\end{figure}

\begin{figure}
\centering
\includegraphics[width = 0.97\columnwidth]{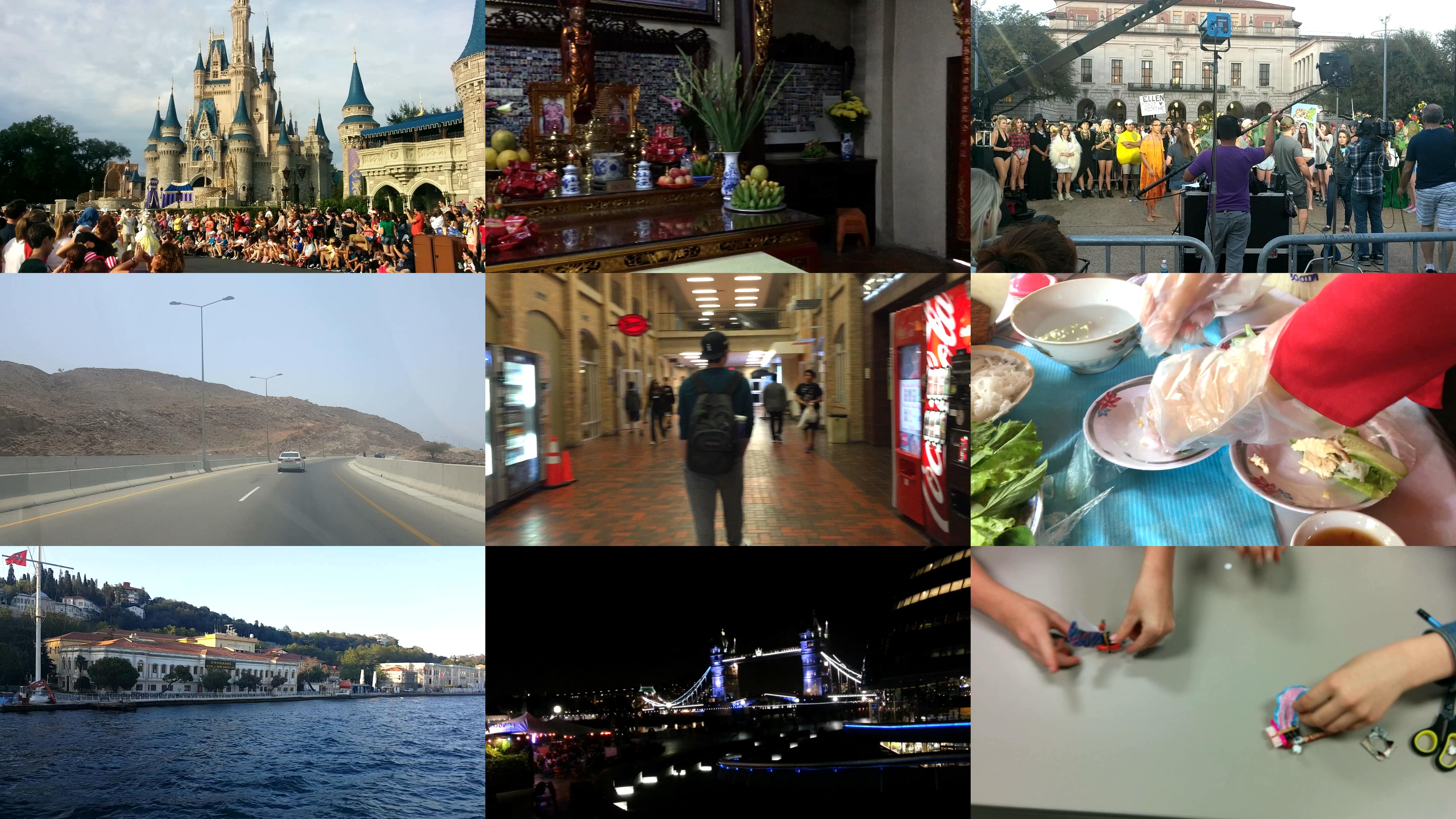}
\caption{Example reference videos from the new LIVE Wild Compressed Video Database.}
\label{screenshot_video_content}
\end{figure}

Since the LIVE VQC database provides subjective scores, specifically Mean Opinion Score (MOS), we randomly sampled the 55 reference videos to match the MOS distribution of the 110 source videos. 
Fig. \ref{MOS_distribution_VQC} shows that the 55 selected reference videos have a very similar MOS distribution as the entire collection of 110 videos LIVE VQC videos. 
Fig. \ref{screenshot_video_content} shows a few example reference videos from the new database. 

\begin{figure}
 \centering
 \subfigure[]{
   \label{NIQE_distribution_VQC:a}
   \includegraphics[width=0.48\columnwidth]{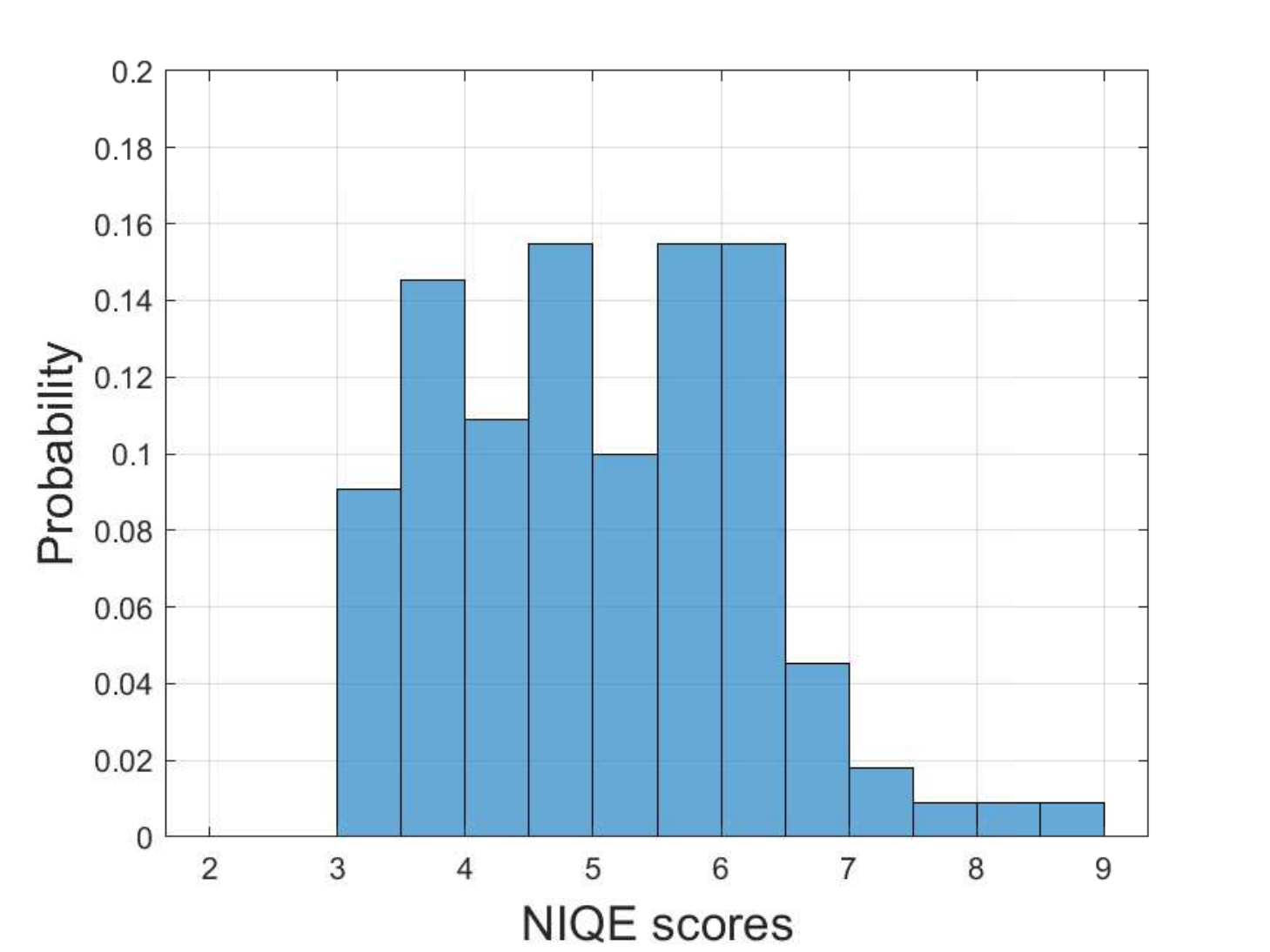}}
 \subfigure[]{
   \label{NIQE_distribution_VQC:b} 
   \includegraphics[width=0.48\columnwidth]{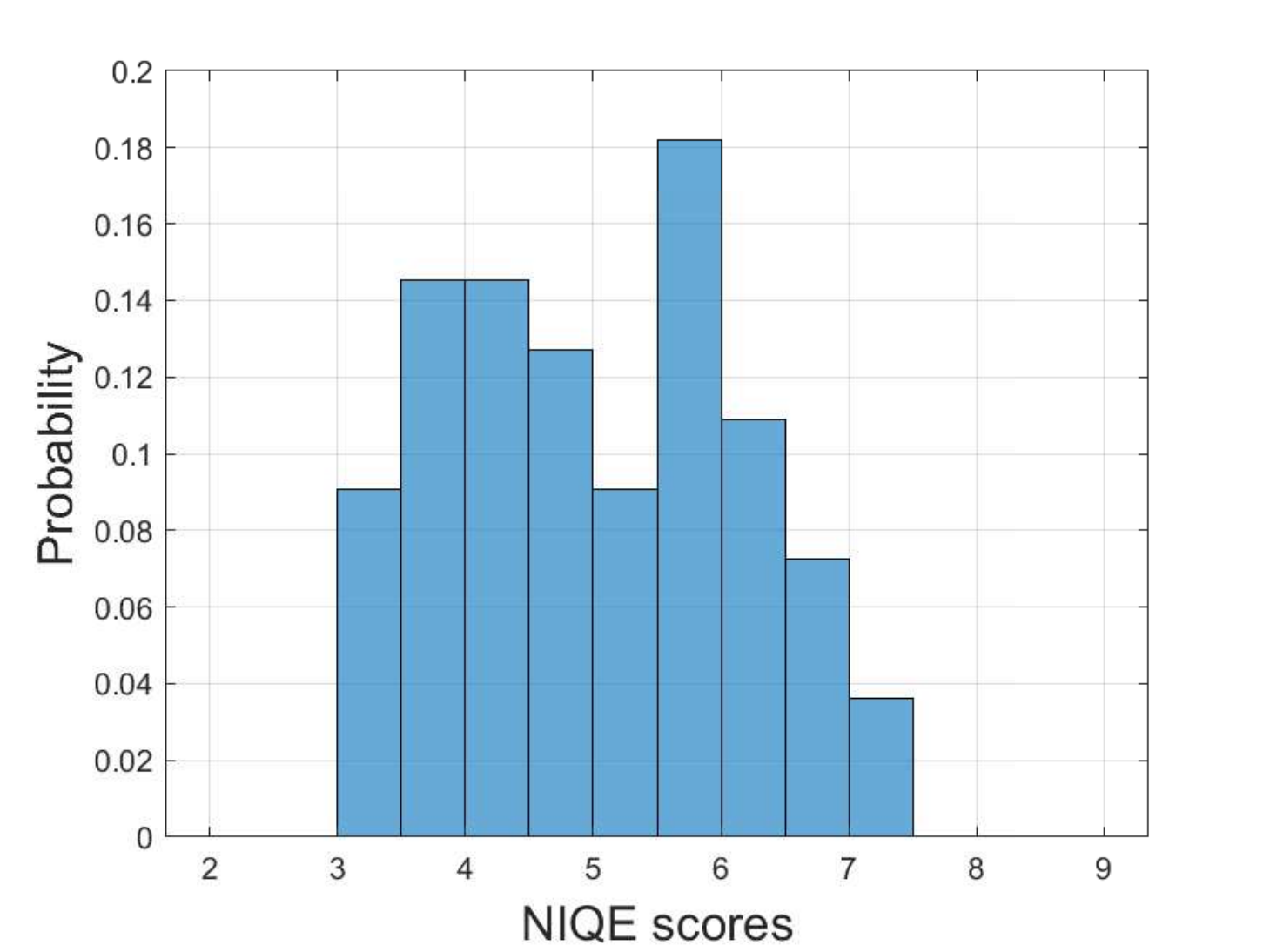}}
 \caption{(a) NIQE score distribution of the 110 source videos in the LIVE Video Quality Challenge Database. The SROCC between the NIQE scores and MOS of the 110 source videos is 0.4989. (b) NIQE score distribution of the 55 selected reference videos in the new LIVE Wild Compressed Video Database. The SROCC between the NIQE scores and MOS of the 55 selected reference videos is 0.5047.}
 \label{NIQE_distribution_VQC} 
\end{figure}

Fig. \ref{NIQE_distribution_VQC} shows the NIQE score distribution of the 110 source videos from the LIVE VQC database as well as the 55 selected reference videos. 
The Spearman’s Rank Ordered Correlation Coefficient (SROCC) between the NIQE scores and MOS of the 55 selected reference videos is also very similar to that of the SROCC correlation between the NIQE scores and MOS of the 110 source videos. 

\begin{figure}
 \centering
 \subfigure[]{
   \label{SI_TI:a}
   \includegraphics[width=0.48\columnwidth]{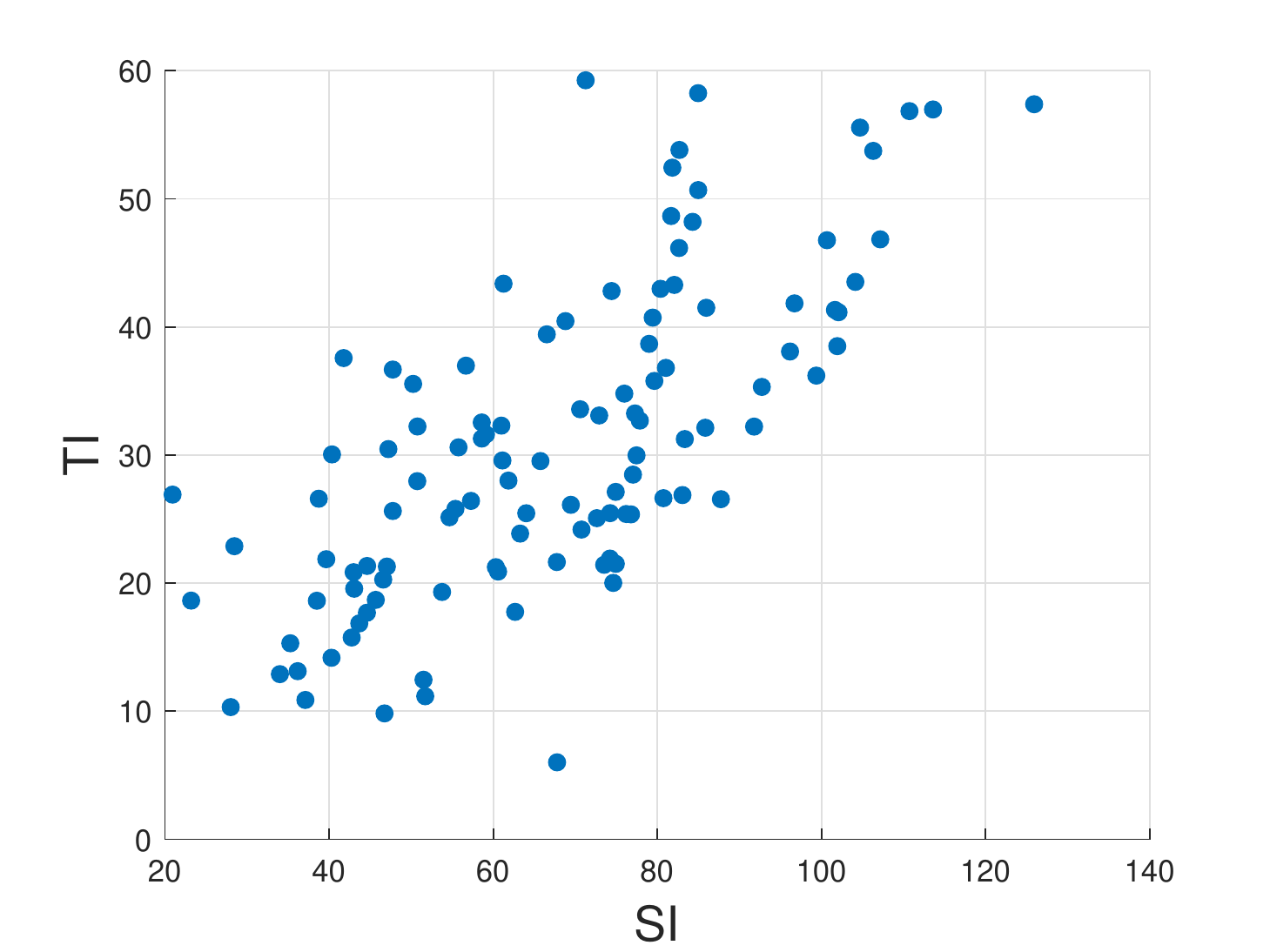}}
 \subfigure[]{
   \label{SI_TI:b} 
   \includegraphics[width=0.48\columnwidth]{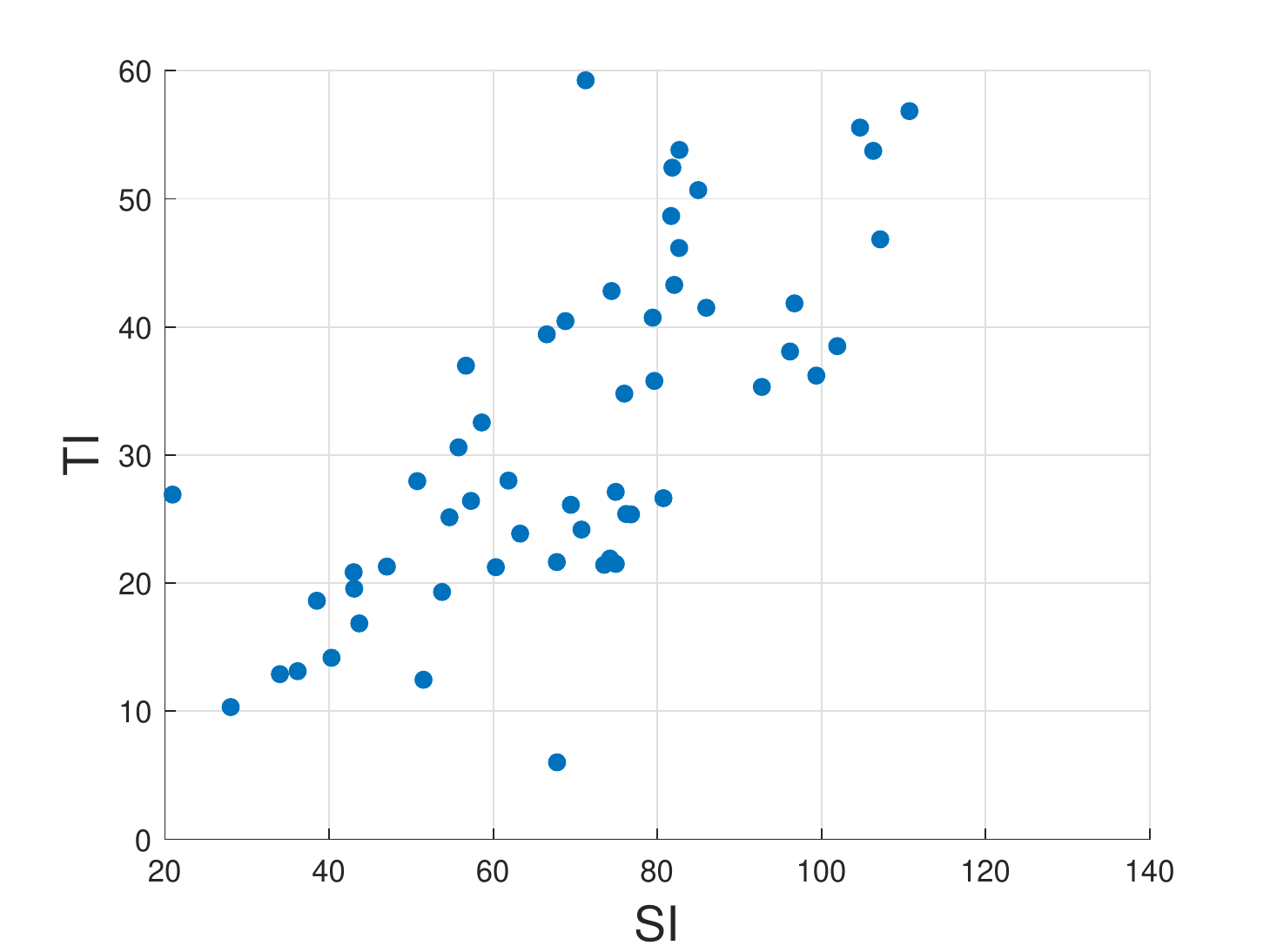}}
 \caption{(a) SI against TI of the 110 source videos in the LIVE Video Quality Challenge Database. (b) SI against TI of the 55 selected reference videos in the new LIVE Wild Compressed Video Database.}
 \label{SI_TI} 
\end{figure}

As a further measure of similarity, we also computed the Spatial Information (SI) and Temporal Information (TI) \cite{series2012methodology} on all of the videos. 
These quantities roughly measure the spatial and temporal richness and variety of the video contents. 
SI and TI are defined as follows:

\begin{equation}
SI = \textrm{max}_{\mathit{time}}\left \{ std_{space}\left [ \textit{Sobel}( F_{n}(i, j)) \right ] \right \},
\label{SI}
\end{equation}

\begin{equation}
TI = \textrm{max}_{\mathit{time}}\left \{ std_{space}\left [  M_{n}(i, j) \right ] \right \},
\label{TI}
\end{equation}

\noindent
where $F_{n}$ denotes the luminance component of a video frame at instant $n$, $(i, j)$ denotes spatial coordinates within this frame, and $M_{n} = F_{n} - F_{n+1}$ is the frame difference operation. 
A Sobel filter is denoted as $Sobel(F_{n})$. 
Fig. \ref{SI_TI} shows that the video contents we selected widely span the SI-TI space, and have a similar SI-TI distribution as the source videos in the LIVE VQC database. 

\subsection{Compressed Video Generation}
\begin{figure}
\centering
\includegraphics[width = 1\columnwidth]{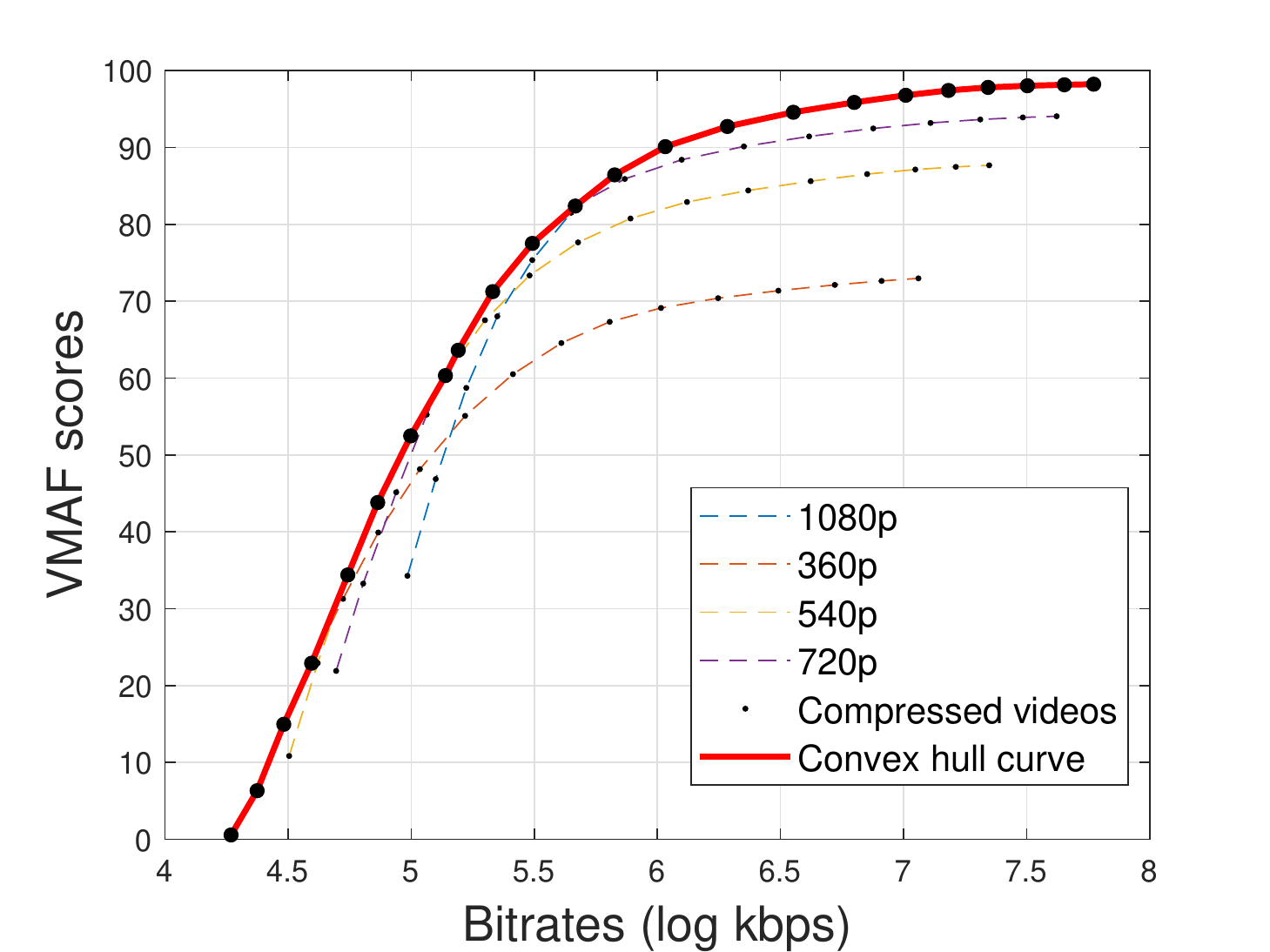}
\caption{Convex hull curve used to select compressed videos of an example content.}
\label{convext_curve}
\end{figure}

Each reference video was subjected to two processing steps to simulate stages that may occur when, for example, a video is uploaded to a social media site. 
First, each video was spatially down-scaled (if needed) to four resolutions: 1080p, 720p, 540p, and 360p. 
Second, we generated these compressed videos at each down-scaled resolution using H.264 at 17 compression levels using CRF: 1, 4, 7, ..., 49. 
Thus, for each video content, there is one 1080p reference video and 68 altered versions of it (four resolutions) x (17 compression levels). 
This yielded a total of 3740 videos with downsample/compression distortions, which is far too many to all be viewed by the limited number of subjects that typically subscribe to a psychometric quality study in a laboratory. 
Therefore, we adopted a strategy similar to that used by video service providers to determine bitrate allocation vs. predicted quality. 
To do this, we calculated the VMAF scores of all the videos along with their bitrates, and used these to plot the convex hull curve of each video content. 
Based on this curve, we selected four videos on the curve and included them in the database. 
An example of the convex hull curve of one video content is shown in Fig. \ref{convext_curve}. 
We chose VMAF scores to draw the convex hull curve because it is a state-of-art FR VQA model that is already used in this way in the streaming industry \cite{aaron2015per}, and its scores map approximately linearly against perceptual quality, which is convenient for video selection and distinction. 
To create downscaled/compressed versions of each content that are perceptually separable from each other, we selected four compressed videos on the curve having VMAF adjacent score differences approximately between 10 and 20. 
The compressed videos at the four compression levels have VMAF scores within the ranges [80,90], [60,70], [40,50], and [20,30], respectively. 
We applied the same process on all contents, yielding four perceptually different downscaled/compressed versions of each, or 220 intentionally impaired UGC videos overall, in addition to the original 55 (275 total). 
A summary of the distributions of video resolutions at each compression level is listed in Table \ref{resolution_dist}. 
Since the convex hull curve of each content is unique, likewise the parameter combinations selected for each content are as well. 

We note that using VMAF to create the compressed content must be regarded as introducing a database bias that VMAF (or models using the same features, like VIF \cite{sheikh2006image} or ST-RRED \cite{soundararajan2012video}) would unfairly benefit from, and indeed, this bias has been observed. 
Hence these models are excluded from the later comparisons. 

\begin{table}
\caption{Distribution of Video Resolutions at Each Compression Level}
 \centering
\label{resolution_dist}
\begin{tabular}{|c|c|c|c|c|}
\hline
\diagbox{Resolution}{Compression level} & 1  & 2  & 3  & 4  \\ \hline
1080p                                   & 11 & 0  & 0  & 0  \\ \hline
720p                                    & 33 & 17 & 1  & 0  \\ \hline
540p                                    & 11 & 33 & 26 & 6  \\ \hline
360p                                    & 0  & 5  & 28 & 49 \\ \hline
\end{tabular}
\end{table}

\subsection{Human Study}
In order to obtain subjective quality labels on each video (for model training, testing, and comparison), we conducted a human study in the LIVE subjective study laboratory. 
The participating subjects were mostly UT-Austin students without a background in video quality. 
Each subject completed two test sessions of duration about 45 mins, separated by at least 24 hours. 
The database of 275 videos was divided randomly into two parts in each session, one containing 27 contents and another containing 28 contents, including both reference videos and their respective four downscaled/compressed versions, hence each subject viewed 135 and 140 videos in consecutive sessions on different days. 
The videos were played in a randomized order with each video shown only once during each session, and where different distorted versions of each unique content were separated by at least 5 videos. 
The source videos were included as references. 
The total number of subjects that took part in the study was 40, and all of them successfully finished both sessions. 

\begin{figure}
 \centering
 \subfigure[]{
   \label{rating_bar}
   \includegraphics[width=0.47\columnwidth]{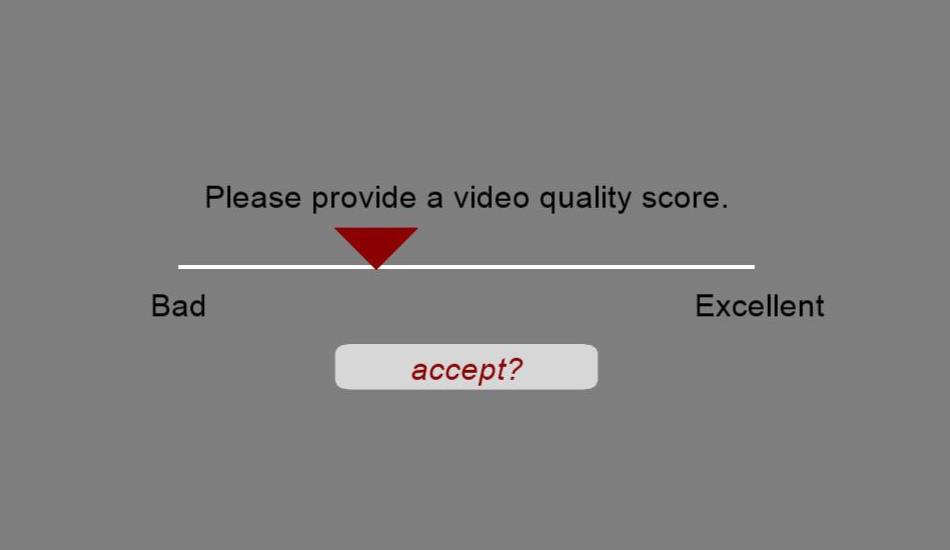}}
 \subfigure[]{
   \label{interface_instruction} 
   \includegraphics[width=0.47\columnwidth]{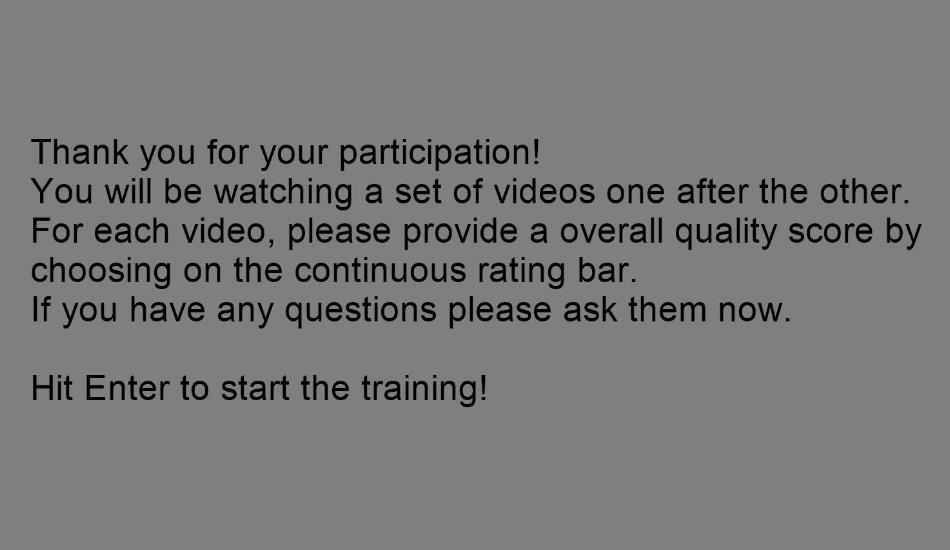}}
 \caption{(a) Screenshot of the rating bar shown to the subjects. (b) Screenshot of the instruction interface shown to the subjects.}
 \label{interface} 
\end{figure}

At the start of the first session, each subject participated in a visual acuity (Snellen) test, and were asked whether they had any uncorrected visual deficiency. 
A viewing distance of approximately two feet was maintained during the test. 
The video sequences were displayed on an HP VH240a 23.8-inch 1080p monitor at their native 1920 x 1080 resolution, and the subjects were asked to adjust the height and angle of the monitor to find their best position. 
Before starting the experiment, each subject was required to read and sign a consent form including general information about the human study, then the procedures and requirements of the test were explained. 
A short training session was also presented at the beginning of the first session, using a different set of videos than in the test experiment, to help the subjects become familiar with the procedures. 
After watching each video, each subject was asked to provide an overall opinion score of the video's quality by dragging a slider along a continuous rating bar. 
As shown in Fig. \ref{rating_bar}, the quality range was labeled from low to high with two adjectives: Bad and Excellent. 
The subjective scores obtained from the subjects were converted to numerical quality scores in [1, 100]. 
A screenshot of the subjective study instruction interface is shown in Fig. \ref{interface_instruction}. 
The interface was developed on a Windows PC using the PsychoPy software \cite{peirce2009generating}. 

\subsection{Data Processing}
The subjective MOS were then computed following the procedures described below. 
The collected raw scores were first converted into Z-scores. 
Let $s_{ijk}$ denote the score rated by the $i$-th subject on the $j$-th video in the session $k = \{1, 2\}$. 
The Z-scores were computed as follows: 

\begin{equation}
z_{ijk} = \frac{s_{ijk} - \bar{s}_{ik}}{\sigma_{ik}},
\label{z-score}
\end{equation}

\noindent 
where $\bar{s}_{ik}$ is the mean of the collected raw scores over all videos assessed by subject \textit{i} in session \textit{k}, and $\sigma_{ik}$ is the standard deviation.

\subsection{Analysis}
We first conducted a consistency test. 
We randomly split the subjective ratings gathered on each video into two disjoint equal groups, and computed and compared the MOS on each video. 
We repeated the random splits 1000 times and the median SROCC between the two groups was \textbf{0.9849}, which is excellent. 
Since the inter-subject agreement was so high, we elected not to apply a subject rejection process. 
We also compared the correlation of the MOS of the 55 reference videos used in our new study, against the previously obtained (crowdsourced) MOS on the same 55 videos from the LIVE VQC database, and found SROCC = \textbf{0.8390}. 
The SROCC results show that our data are reliable. 

\begin{figure}
 \centering
 \subfigure[]{
   \label{MOS_Compressed:a}
   \includegraphics[width=0.48\columnwidth]{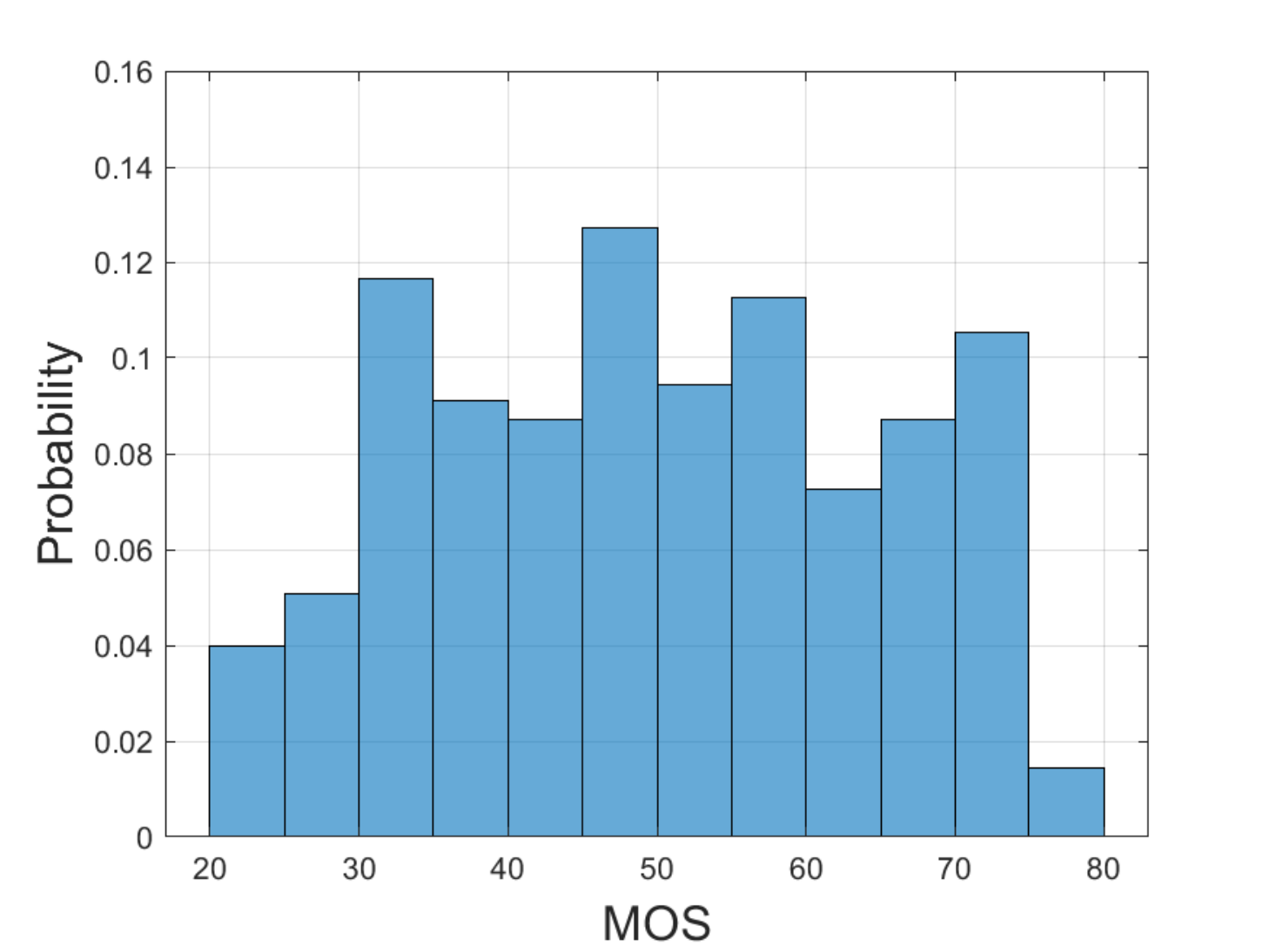}}
 \subfigure[]{
   \label{MOS_Compressed:b} 
   \includegraphics[width=0.48\columnwidth]{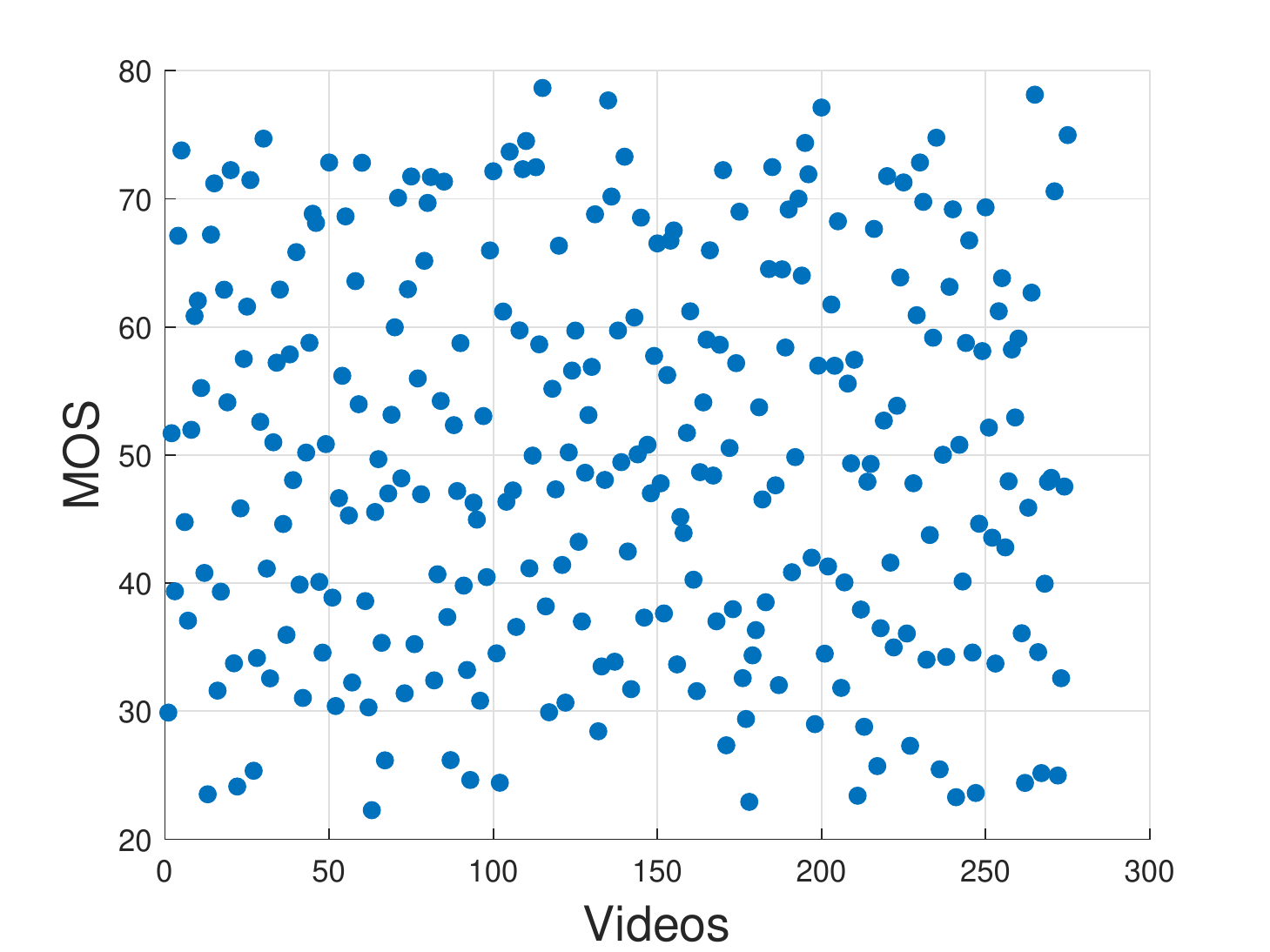}}
 \caption{(a) MOS distribution across the entire LIVE Wild Compressed Video Quality Database. (b) Scatter plot of the MOS obtained on all the videos in the database.}
 \label{MOS_Compressed} 
\end{figure}

\begin{figure}
\centering
\includegraphics[width = 1\columnwidth]{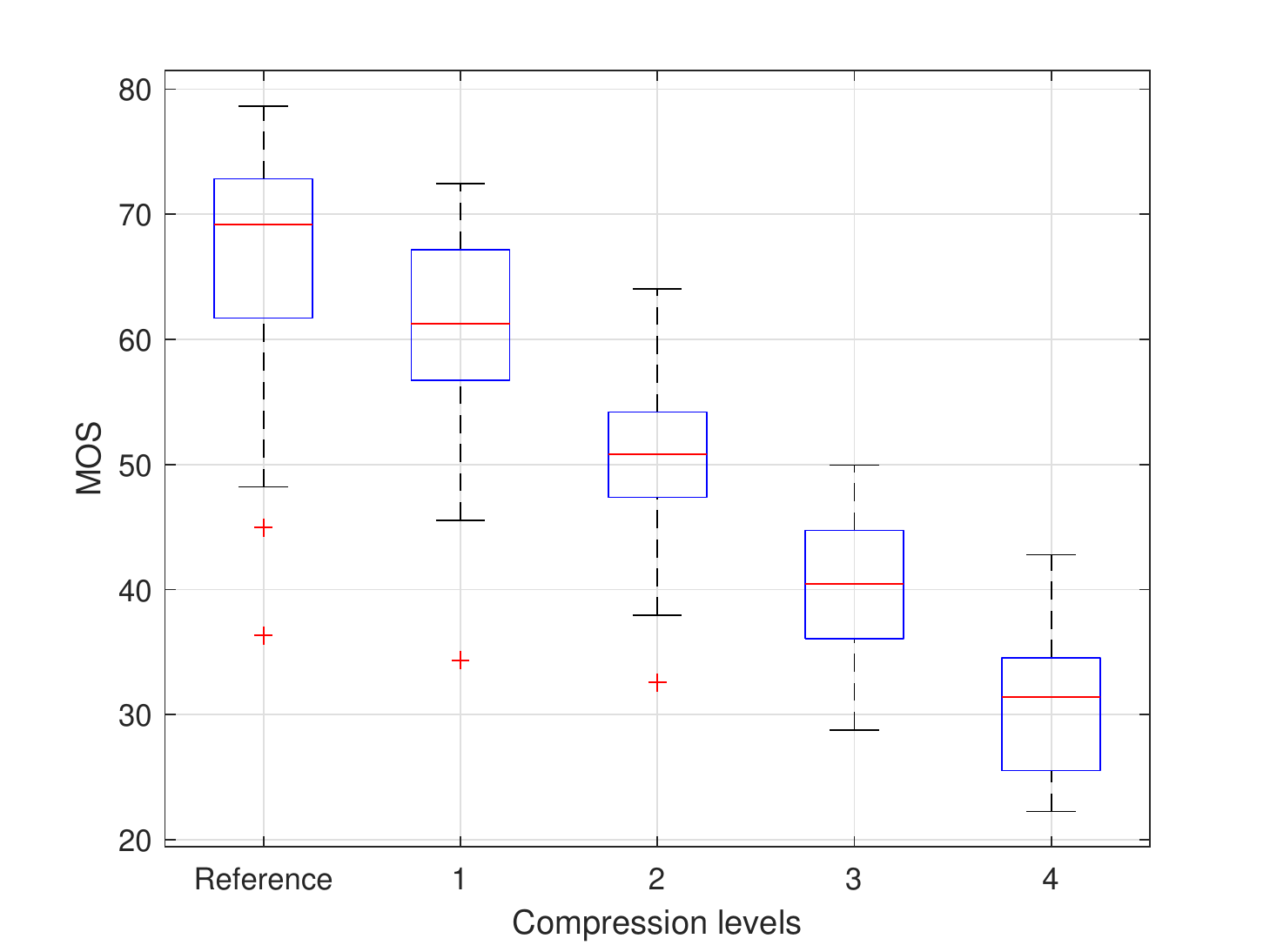}
\caption{Box plot of MOS of videos in the LIVE Wild Compressed Video Quality Database for different compression levels. The central red mark represents the median, while the bottom and top edges of the box indicate the 25th and 75th percentiles, respectively. The whiskers extend to the most extreme data points not considered outliers, while outliers are plotted individually using the '+' symbol.}
\label{boxplot_MOS_comlevel}
\end{figure}

\begin{figure}
\centering
\includegraphics[width = 1\columnwidth]{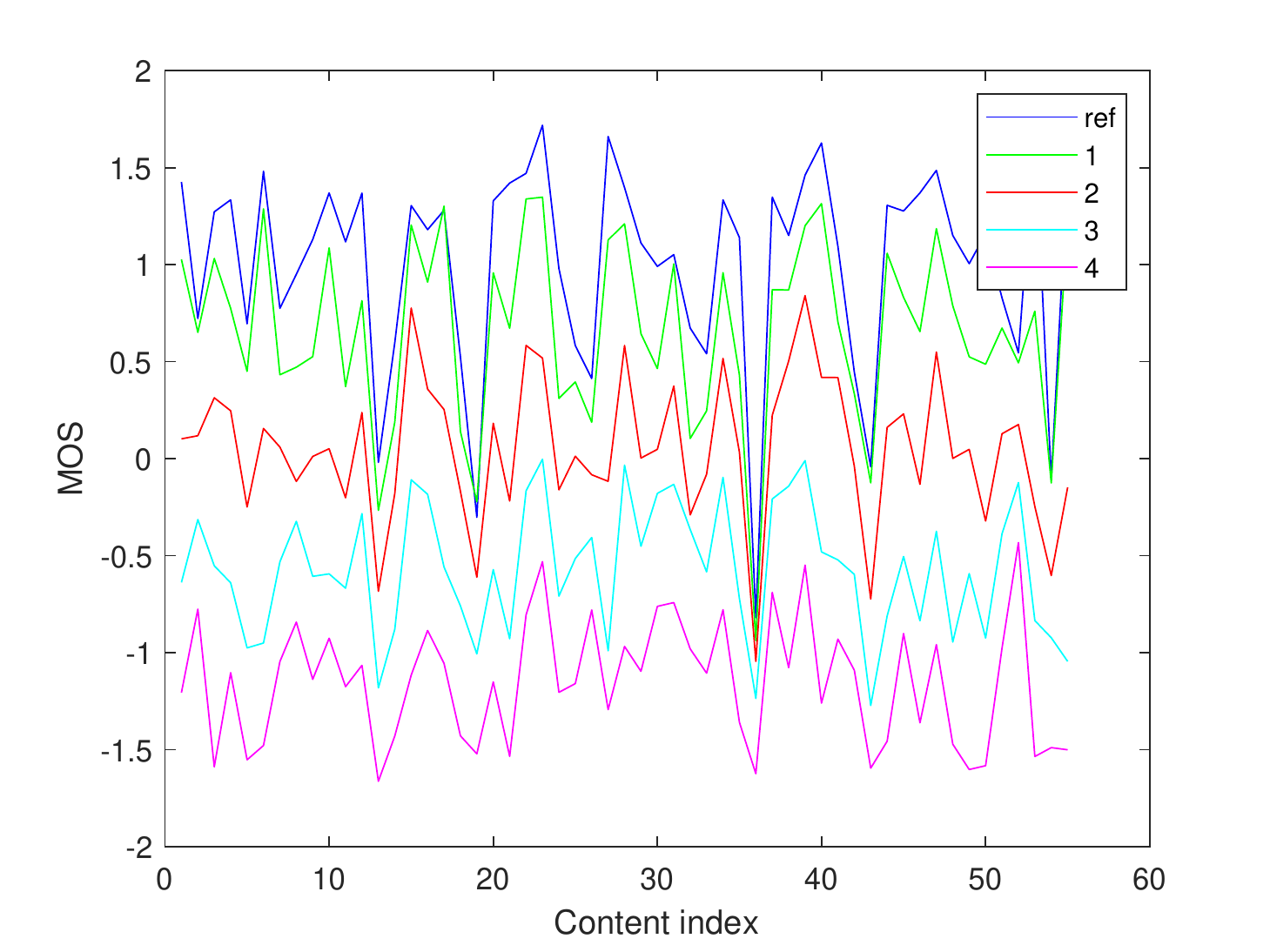}
\caption{MOS of all contents for four different compression (distortion) levels and reference videos, coded by color.}
\label{curve_plot_MOS_trend}
\end{figure}

The overall MOS distribution and a scatter plot of the MOS of the LIVE Wild Compressed Video Quality Database are plotted in Fig. \ref{MOS_Compressed}. 
Fig. \ref{boxplot_MOS_comlevel} is a box plot of the MOS of the reference videos and the four compression levels (but across downscaling). 
The MOS decreases as the compression is increased. 
Fig. \ref{curve_plot_MOS_trend} plots the MOS across all contents, each color coded at a different compression level. 
While the curves are nicely separated by content, it is important to observe the mixing of MOS across contents, which is largely caused by the reference distortions, rather than the applied downscaling/compression. 

\section{1stepVQA Model}
\label{1stepVQA_model}

\begin{figure}
 \centering
 \subfigure[]{
   \label{Overview:main}
   \includegraphics[width=0.95\columnwidth]{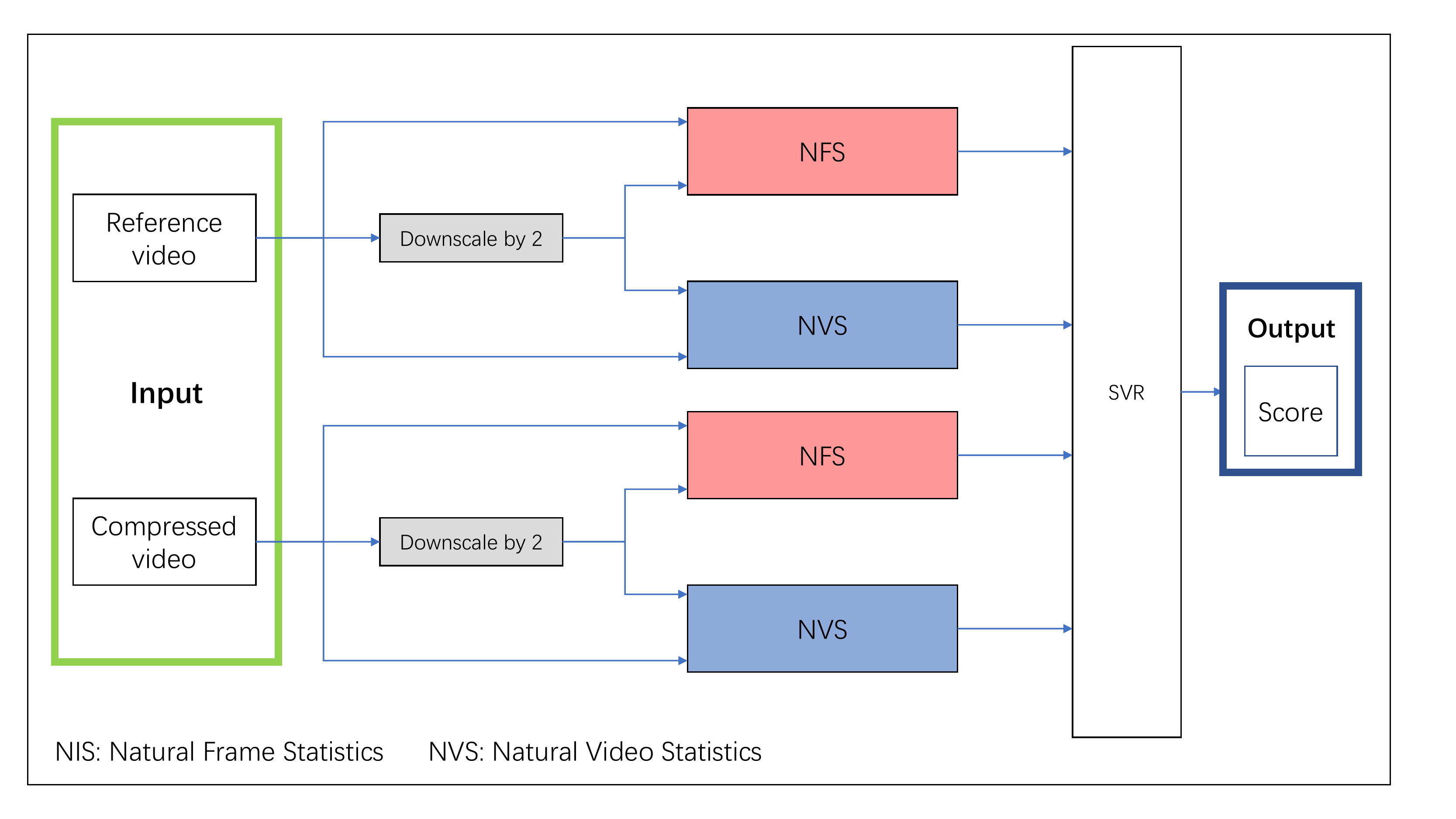}}
 \subfigure[]{
   \label{Overview:NFS}
   \includegraphics[width=0.8\columnwidth]{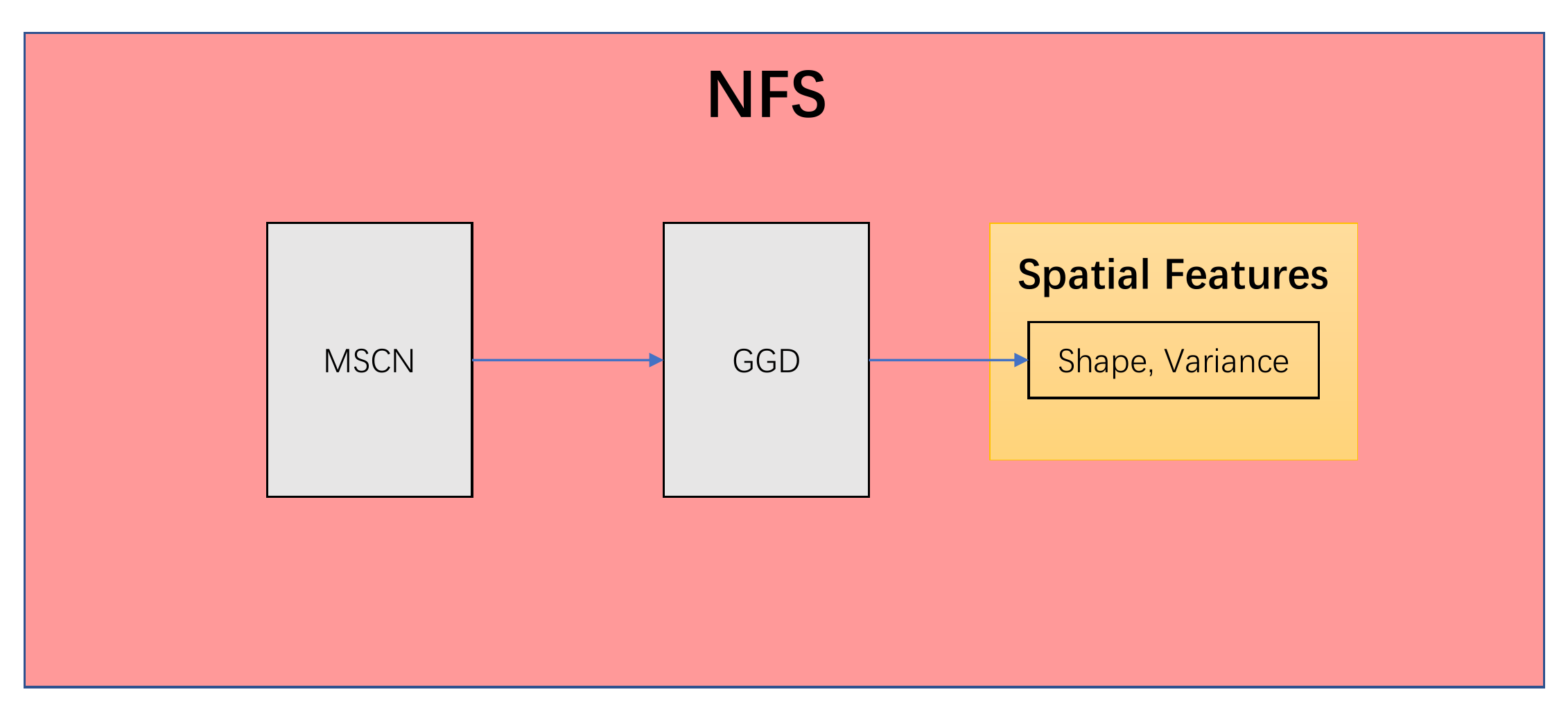}}
 \subfigure[]{
   \label{Overview:NVS} 
   \includegraphics[width=0.8\columnwidth]{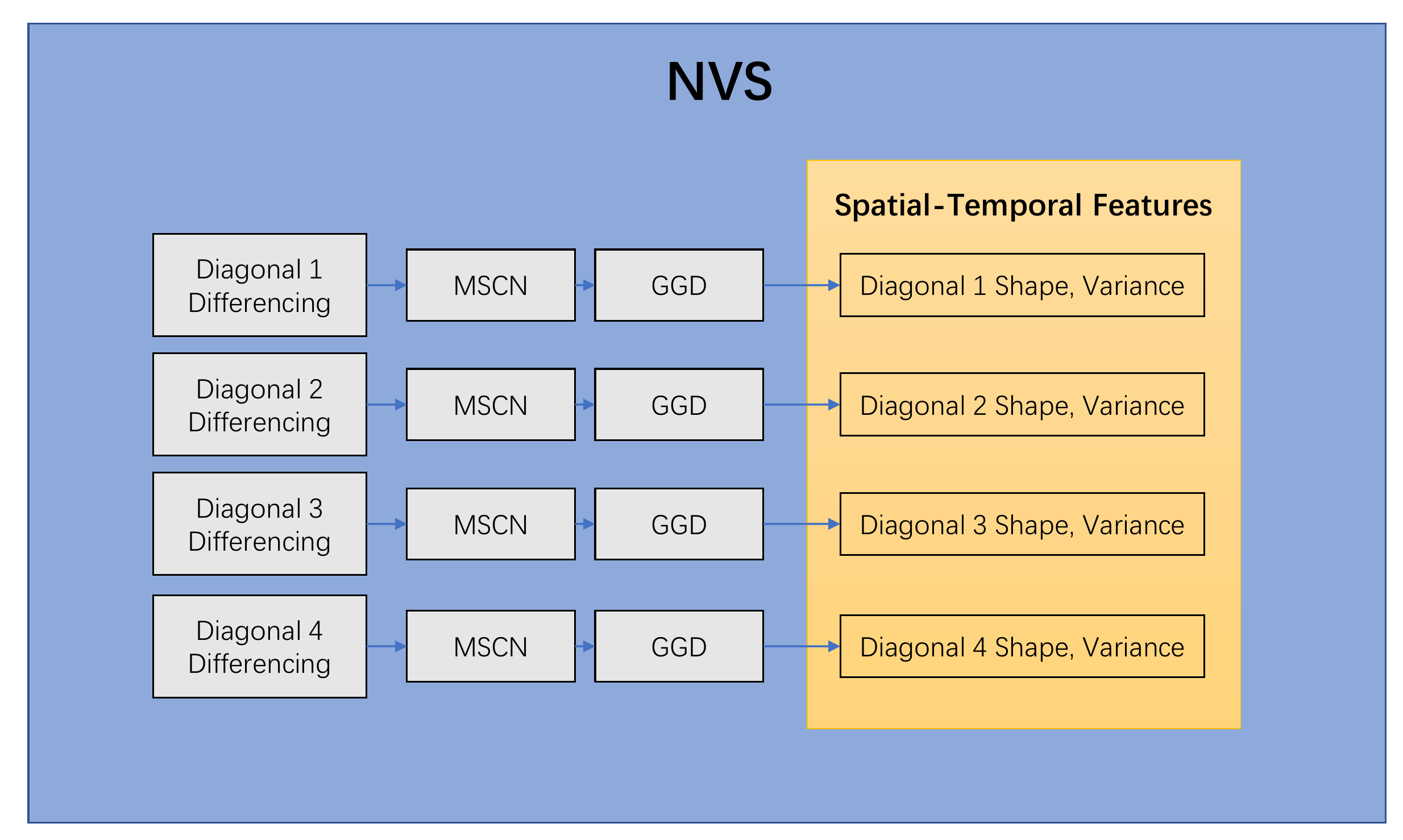}}
 \caption{(a) 1stepVQA overview. (b) Natural Frame Statistic (NFS) processing flow. (c) Natural Video Statistic (NVS) processing flow.}
 \label{Overview} 
\end{figure}

Research on the relationship between visual perception and the statistics of natural images has aroused widespread attention \cite{wandell1995foundations}. 
These models become well-behaved when expressed in bandpass domain, e.g., wavelets, DCT and Gabor transforms, etc \cite{srivastava2003advances}. 
The 1stepVQA model utilizes an extended set of video NSS models, that capture directional bandpass space-time attributes and uses them to predict quality. 
An overview of the 1stepVQA model is shown in Fig. \ref{Overview}. 
We discuss the spatial features and spatial-temporal features next. 

\subsection{Natural Frame Statistics}
Here forward to distinguish them from more general spatial-temporal NSS models, we shall refer to spatial (frame) NSS as Natural Frame Statistics (NFS). 
The application of NFS models in 1stepVQA is quite similar to BRISQUE \cite{mittal2012no}. 
To capture quality-aware NFS features, apply a local non-linear bandpass operation on the luminance component of each video frame, in the form of local mean subtraction followed by divisive normalization. 
Given a video having $T$ luminance frames {$I_{1}, I_{2}..., I_{t}..., I_{T}$}, define spatial coordinates $(i,j)$, where $i \in 1,2...M, j \in 1,2...N$ are spatial indices of $I_{t}$, then the mean subtracted contrast normalized coefficients (MSCN) $\hat{I}_{t}$ are: 

\begin{equation}
\begin{aligned}
&\hat{I}_{t}(i,j) = \frac{I_{t}(i,j) - \mu_{t}(i,j)}{\sigma_{t}(i,j) + C}\\
\label{MSCN}
\end{aligned}
\end{equation}

where

\begin{equation}
\begin{aligned}
&\mu_{t}(i,j) = \sum_{k = -K}^{K}\sum_{l = -L}^{L}w _{k,l}I_{t}(i-k,j-l)\\
\label{MSCN_mean}
\end{aligned}
\end{equation}

and

\begin{equation}
\begin{aligned}
&\sigma_{t}(i,j) = \sqrt{\sum_{k = -K}^{K}\sum_{l = -L}^{L}w _{k,l}(I_{t}(i-k,j-l) - \mu_{t}(i,j))^{2}}
\label{MSCN_var}
\end{aligned}
\end{equation}

\noindent
where $C = 1$ is a stability constant, and $w = \{w_{k,l}|k = -K,...,K, l = -L,...,L\}$ is a 2D circularly-symmetric Gaussian weighting function with $K = L = 3$. 

\begin{figure}
\centering
\includegraphics[width = 1\columnwidth]{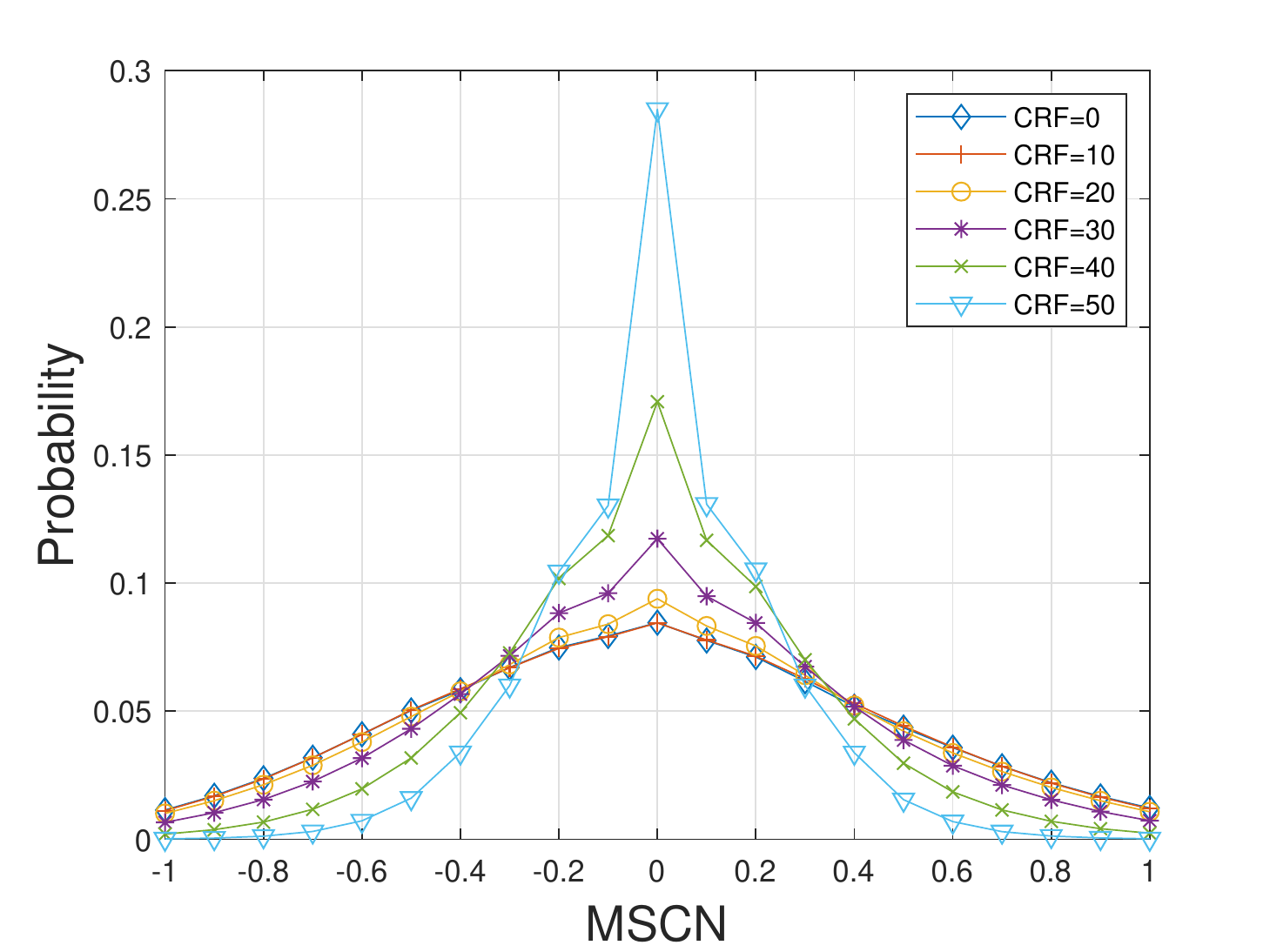}
\caption{Histograms of MSCN coefficients for a frame of a natural undistorted video and various H.264 compressed versions of it in different CRFs.}
\label{Aerial_MSCN_spatial}
\end{figure}

If there are spatial distortions present, then the statistical distribution of frame MSCN coefficients tend to become predictably altered \cite{mittal2012no}. 
For example, Fig. \ref{Aerial_MSCN_spatial} plots the histograms of the MSCN coefficients of a frame of an undistorted video and various compressed versions of it. 

Following \cite{mittal2012no, mittal2013making}, we use a Generalized Gaussian Distribution (GGD) model of the MSCN coefficients. 
A GGD with zero mean is given by: 

\begin{equation}
\begin{aligned}
f(x;\alpha ,\sigma ^{2})=\frac{\alpha }{2\beta \Gamma (1/\alpha)}exp(-(\frac{|x|}{\beta })^{\alpha })
\label{GGD}
\end{aligned}
\end{equation}

where

\begin{equation}
\begin{aligned}
\beta =\sigma \sqrt{\frac{\Gamma (1/\alpha)}{\Gamma(3/\alpha)}}
\label{GGD_beta}
\end{aligned}
\end{equation}

and $\Gamma(\cdot)$ is the gamma function:

\begin{equation}
\begin{aligned}
\Gamma (\alpha)=\int_{0}^{\infty}t^{\alpha -1}e^{-t}dt   \quad  a>0.
\label{GGD_gamma}
\end{aligned}
\end{equation}

There are two GGD parameters: a shape parameter $\alpha$, and a spread parameter $\sigma$. 
These are estimated using the popular moment-matching approach in \cite{sharifi1995estimation}. 
Since we have found that the spread parameter $\sigma$ of the GGD does not contribute to better quality prediction performance, only the shape parameters $\alpha$ is included as part of the feature set. 

\subsection{Natural Video Statistics}
Here we refer to the spatial-temporal NSS of videos as Natural Video Statistics (NVS), which are also significantly affected by distortions, such as jitter, ghosting, and motion compensation mismatches, as well as compression and transmission artifacts. 
Moreover, UGC videos are often afflicted by mixed in-capture distortions, such as camera shake, under- or over-exposure, sensor noise, and color distortions, which cannot be easily modeled. 

\begin{figure}
\centering
\includegraphics[width = 1\columnwidth]{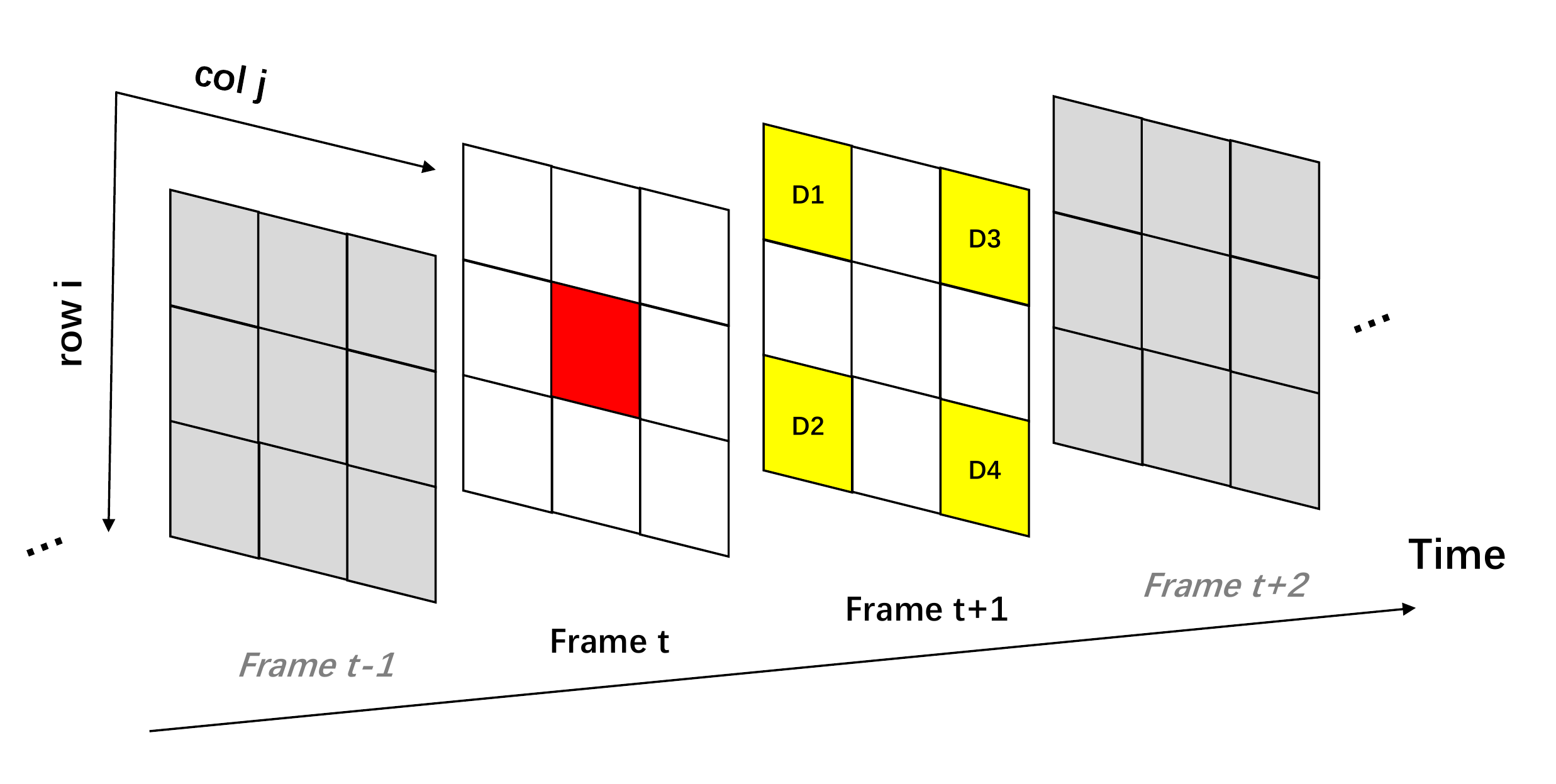}
\caption{Depiction of a pixel in frame $t$ and the four displaced pixels in frame $t + 1$ it is differenced with.}
\label{diag_frm_diff}
\end{figure}

While it is difficult to find regularities in the statistics of motion, there are strong regularities of temporal bandpass videos, such as frame difference signals \cite{soundararajan2012video}. 
However, our model goes further, and also utilizes models of the statistics of displaced frame differences. 
Given two adjacent frames, perform a spatial translation of one of them, then compute the displaced frame difference between the two frames. 
By doing so, we seek to capture space-time statistics of videos without computing motion. 
As it turns out, bandpass processing of displaced frame differences are also very regular, and predictive of quality. 
For simplicity, we used four diagonal directions to compute displaced frame differences. 
Given a video with $T$ luminance frames {$I_{1}, I_{2}..., I_{t}..., I_{T}$} and spatial coordinates $(i,j)$, $i \in 1,2...M, j \in 1,2...N$, the four diagonal displaced frame differences between each pair of adjacent shifted frames are defined and depicted in Fig. \ref{diag_frm_diff}. 
The four displaced frame differences are: 

\begin{equation}
\begin{aligned}
D_{t1}(i,j)=I_{t}(i,j)-I_{t+1}(i-1,j-1)
\label{diag_FD1}
\end{aligned}
\end{equation}

\begin{equation}
\begin{aligned}
D_{t2}(i,j)=I_{t}(i,j)-I_{t+1}(i+1,j-1)
\label{diag_FD2}
\end{aligned}
\end{equation}

\begin{equation}
\begin{aligned}
D_{t3}(i,j)=I_{t}(i,j)-I_{t+1}(i-1,j+1)
\label{diag_FD3}
\end{aligned}
\end{equation}

\begin{equation}
\begin{aligned}
D_{t4}(i,j)=I_{t}(i,j)-I_{t+1}(i+1,j+1)
\label{diag_FD4}
\end{aligned}
\end{equation}

The reason that we capture NVS in opposite directions is because they provide complementary, and not redundant distortion information relative to the direction of any motion field. 
One of two opposite directions will generally be more in the direction of local motion, and the other less so, presenting different (more vs. less) statistical regularity in the corresponding frame differences, which can be affected by distortion. 
Then compute the MSCN coefficients of each of the four diagonal displaced frame differences, $\hat{D}_{tk}(i,j)$, $k = 1,...,4$, using (\ref{MSCN}). 
The corresponding values of $\mu_{tk}(i,j)$ and $\sigma_{tk}(i,j)$ are computed in the same way as in (\ref{MSCN_mean}) and (\ref{MSCN_var}). 

\begin{figure*}
 \centering
 \subfigure[]{
   \label{Tango_MSCN:video}
   \includegraphics[width=0.6\textwidth]{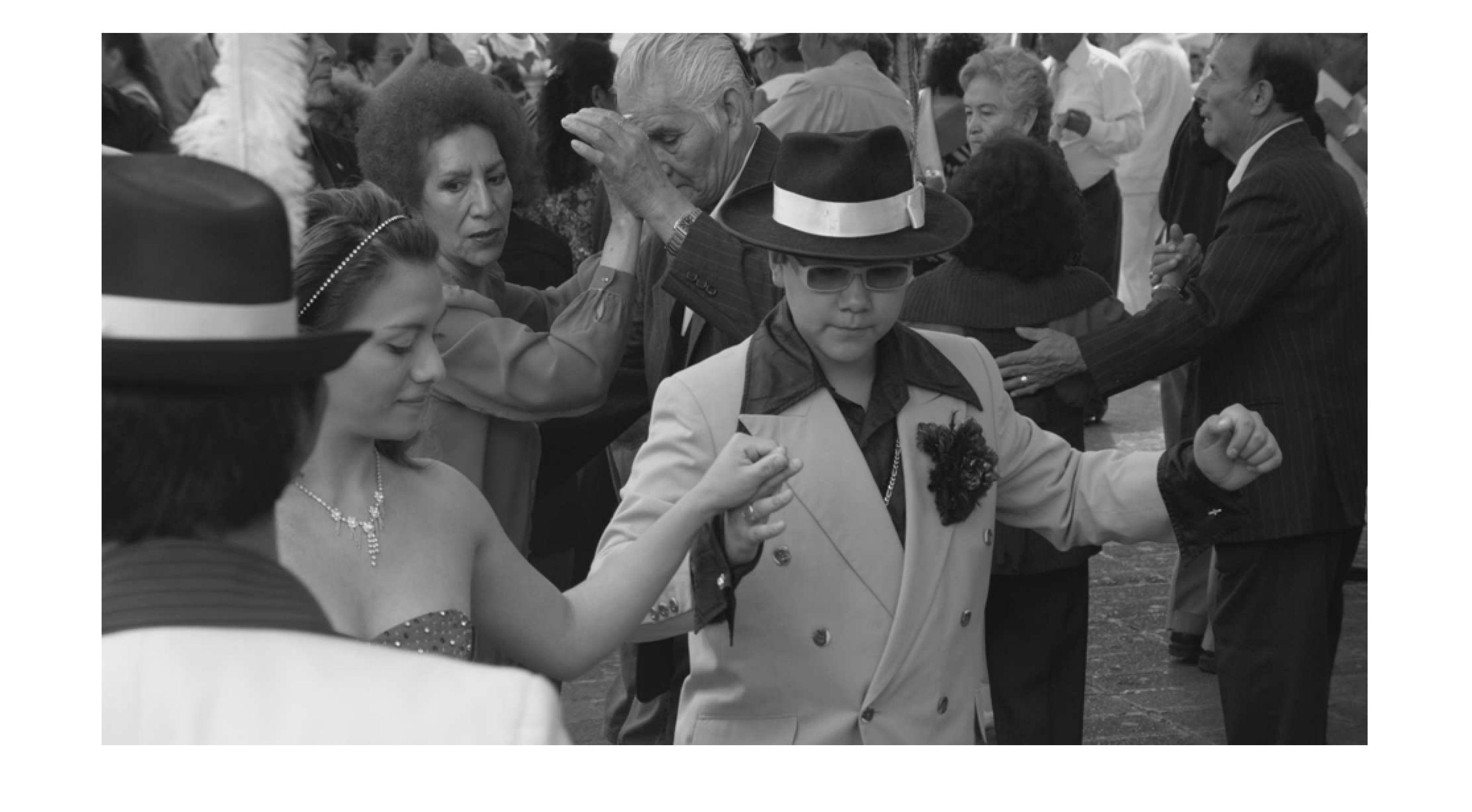}}
 \subfigure[]{
   \label{Tango_MSCN:MSCN_0_0}
   \includegraphics[width=0.44\textwidth]{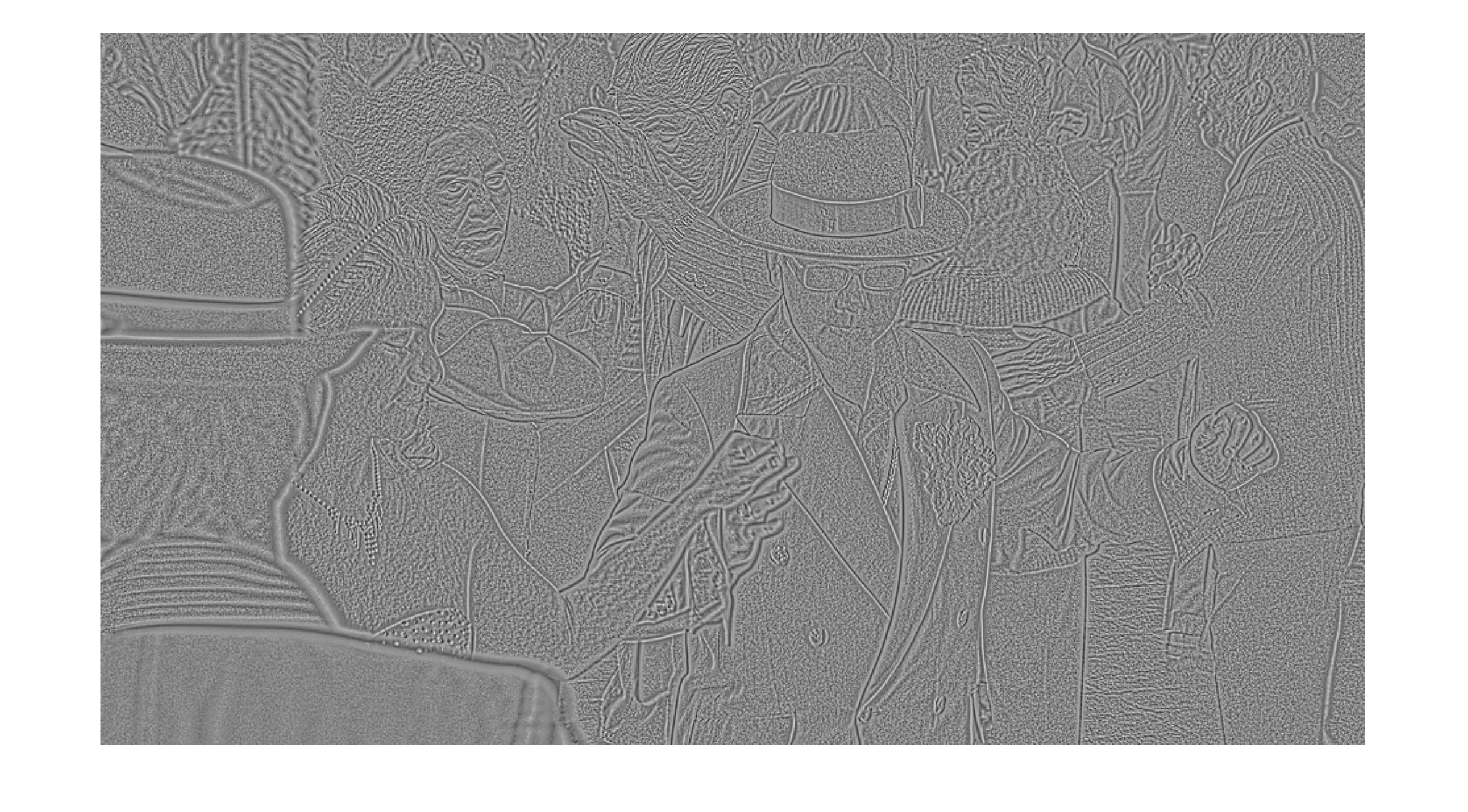}}
 \subfigure[]{
   \label{Tango_MSCN:MSCN_1_1} 
   \includegraphics[width=0.44\textwidth]{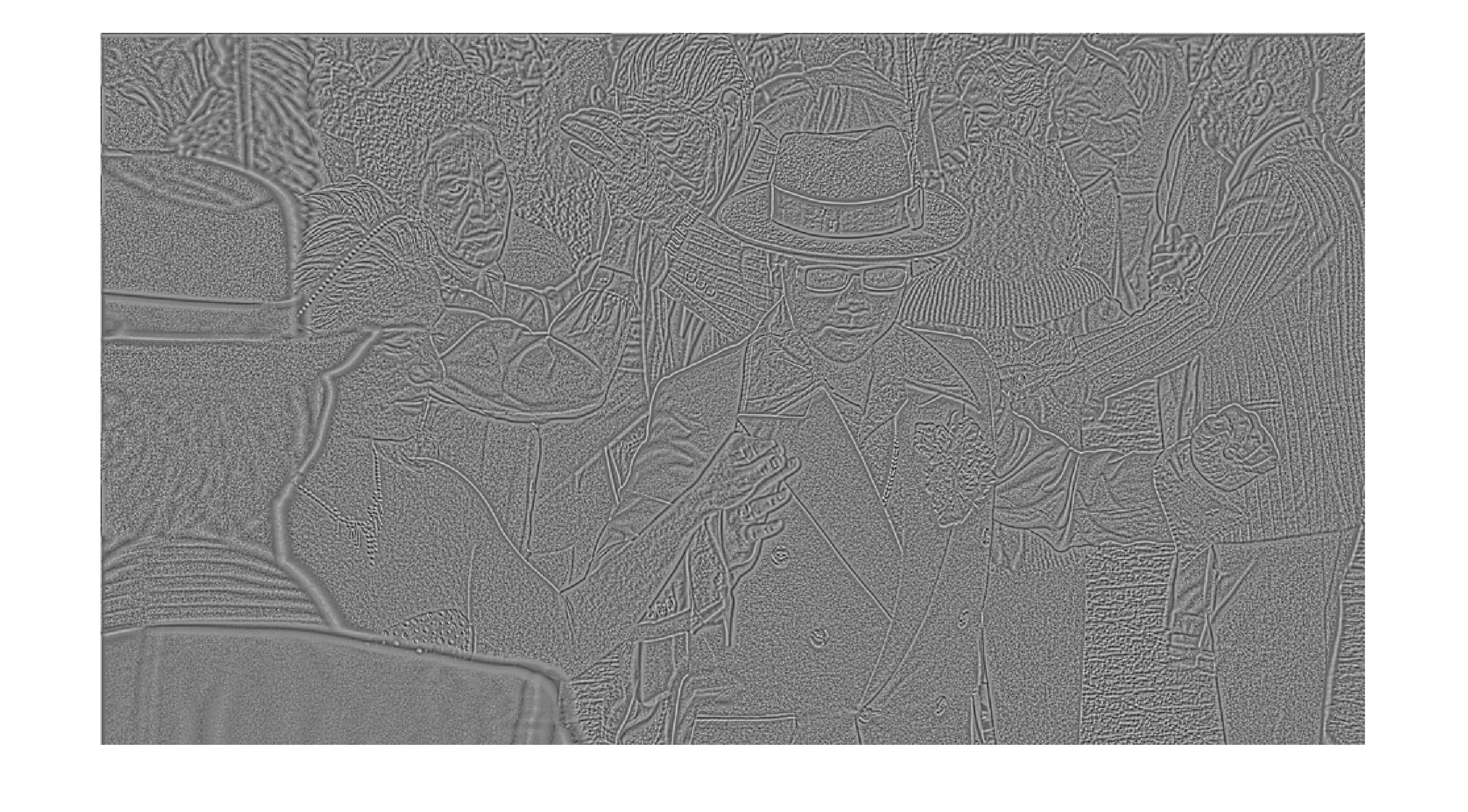}}
 \subfigure[]{
   \label{Tango_MSCN:MSCN_dist_0_0} 
   \includegraphics[width=0.44\textwidth]{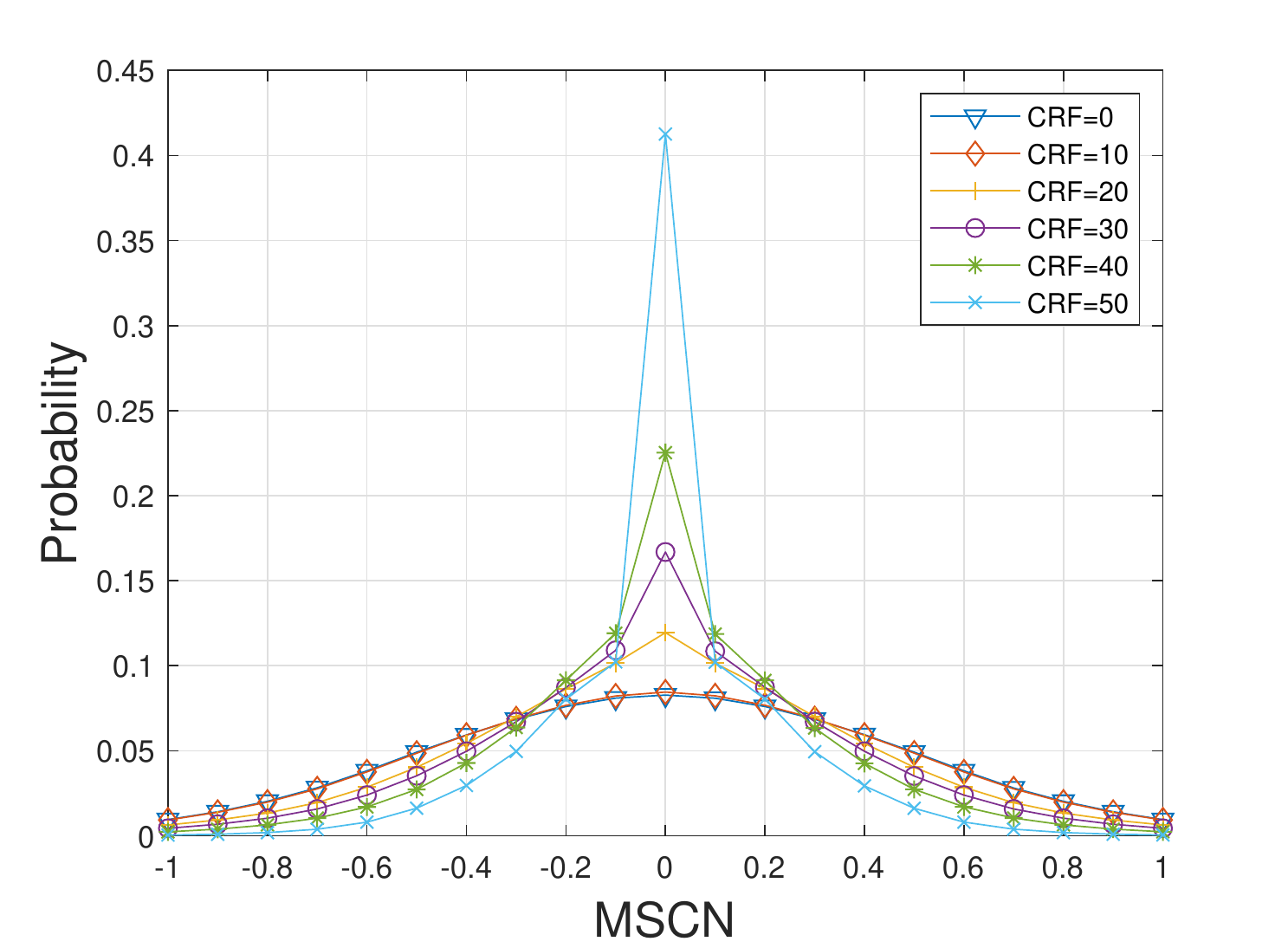}}
 \subfigure[]{
   \label{Tango_MSCN:MSCN_dist_1_1} 
   \includegraphics[width=0.44\textwidth]{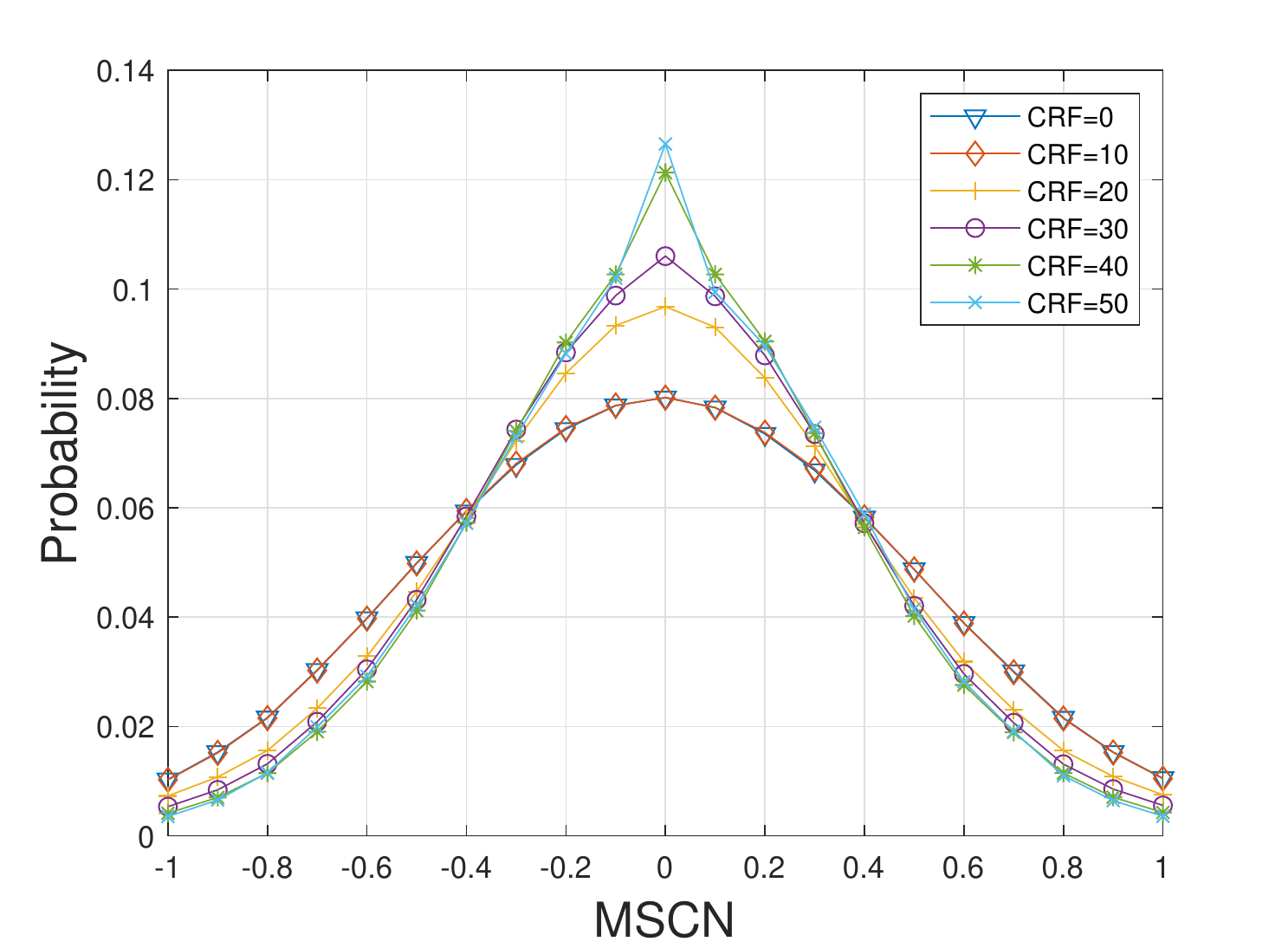}}   
 \caption{ (a) A luminance frame of an undistorted source video. (b) MSCN coefficients of the non-displaced frame difference. (c) MSCN coefficients of the displaced frame difference $\hat{D}_{t1}(i,j)$.  (d) Histogram of MSCN coefficients of the non-displaced frame difference and several compressed versions of it. (e) Histogram of MSCN coefficients of the displaced frame difference $\hat{D}_{t1}(i,j)$ and compressed versions of it.}
\label{Tango_MSCN} 
\end{figure*}

Fig. \ref{Tango_MSCN} shows an examplar high-quality video which we use to illustrate MSCN processing of displaced frame differences, and to compare with the MSCN of non-displaced frame differences. 
Figs. \ref{Tango_MSCN:MSCN_0_0} and \ref{Tango_MSCN:MSCN_1_1} show the MSCN coefficients of a non-displaced frame difference, and a displaced frame difference $\hat{D}_{t1}(i,j)$, respectively. 
As compared with the non-distorted frame difference, which contains no motion information (only change), each displaced frame difference signal reflects directional space-time motion and temporal statistical regularities. 
Figs. \ref{Tango_MSCN:MSCN_dist_0_0} and \ref{Tango_MSCN:MSCN_dist_1_1} plot the histograms of MSCN coefficients of non-displaced and displaced frame differences of the same video, and also of several compressed versions of it. 
While the MSCN histogram of the high-quality source video follows a Gaussian distribution, the histograms of the compressed ones deviate towards a heavy-tailed distribution shape, especially at heavier compressions. 
It is interesting to observe that the histograms of the displaced frame differences are less peaked than the non-displaced frame difference histograms. 

As with NFS, the GGD model (\ref{GGD}) - (\ref{GGD_gamma}) is a good fit to the spatial-temporal MSCN histograms. 
Thus the histograms of each directional difference $\hat{D}_{tk}(i,j)$ are all also well fit with a GGD model, each yielding two GGD parameters. 

However, there are significant differences between the histograms of the non-displaced MSCN frame-difference signals, which are much more peaked than those of the displaced MSCN frame-difference coefficients. 
In other words, the displaced MSCN frame differences tend to retain Gaussianity as the compression parameter is varied. 
Although they remain monotonic and well separated, they are also more regular (and as well we shall see, more predictable). 
One reason for this is that in motion regions where the motion direction is similar to the frame-level displacement, redundancy is better exploited, as around the central subjects' left lapel. 
In directions opposite to motion, displaced MSCN frame differences will produce very marked localized changes, which tend to broaden the MSCN histograms. 
Even on low-motion or static videos, the displaced frame differences become similar to spatial directional gradient operators, which have been previously observed to enhance bandpass statistical regularity and to enhance quality prediction \cite{xue2013gradient,xue2014blind}.

\subsection{Feature Summary and Training}
\label{feature_summary_section}

\begin{table*}
\caption{Summary of 1stepVQA Features}
\label{feature_summary}
\resizebox{\textwidth}{!}{%
\begin{tabular}{|c|c|c|c|c|c|c|c|c|}
\hline
                & \multicolumn{2}{c|}{Scale}  & \multicolumn{2}{c|}{Type}       & \multicolumn{2}{c|}{Source} &                     &                                                      \\ \hline
Feature Index   & 1            & 2            & NFS          & NVS              & Reference    & Compressed   & Feature Description     & Computation Procedure                                \\ \hline
$f_{1}-f_{4}$   & $\checkmark$ & $\checkmark$ & $\checkmark$ &                  & $\checkmark$ & $\checkmark$ & Shape                   & Fit GGD to spatial MSCN coefficients                 \\ \hline
$f_{5}-f_{12}$  & $\checkmark$ & $\checkmark$ &              & $\checkmark$     & $\checkmark$ & $\checkmark$ & Shape, variance         & Fit GGD to D1 displaced difference MSCN coefficients \\ \hline
$f_{13}-f_{20}$ & $\checkmark$ & $\checkmark$ &              & $\checkmark$     & $\checkmark$ & $\checkmark$ & Shape, variance         & Fit GGD to D2 displaced difference MSCN coefficients \\ \hline
$f_{21}-f_{28}$ & $\checkmark$ & $\checkmark$ &              & $\checkmark$     & $\checkmark$ & $\checkmark$ & Shape, variance         & Fit GGD to D3 displaced difference MSCN coefficients \\ \hline
$f_{29}-f_{36}$ & $\checkmark$ & $\checkmark$ &              & $\checkmark$     & $\checkmark$ & $\checkmark$ & Shape, variance         & Fit GGD to D4 displaced difference MSCN coefficients \\ \hline
\end{tabular}%
}
\end{table*}

Table \ref{feature_summary} summarizes the features that comprise 1stepVQA. 
Images and videos are naturally multiscale, and it has been observed and demonstrated that incorporating multiscale information in both space and time (and space-time) enhances quality prediction performance \cite{mittal2012no,wang2003multiscale,soundararajan2012video}. 
We extract all of the listed features in Table \ref{feature_summary}, at two scales. 
Scale 1 is the original video resolution, while scale 2 is the downscaled (by a factor of 2) resolution. 
This is accomplished by Gaussian pre-filtering before down-sampling, as with BRISQUE. 
1stepVQA extracts features from both the source videos and compressed videos. 
While 1stepVQA is strictly a reference VQA model, our feature selection process is highly directed, with bifurcated feature sets designed to capture both NR aspects (of the source videos) and FR aspects (of both). 
A total of 40 1stepVQA features are used, none of which require motion estimation. 

The pooling of features is also simple: 1stepVQA uses average pooling: All features are computed on a per frame basis, then averaged across all frames. 
In our implementation, a Support Vector Machine Regressor (SVR) is used to learn a regression model from the extracted feature space to quality scores. 
Similar approaches are implemented in \cite{mittal2012no,saad2014blind,korhonen2019two}. 

In Section \ref{1stepVQA_intra_section}, we study the performance of 1stepVQA, and compare it against other models and feature combinations.

\section{Performance and Analysis}
\label{performance_and_analysis}
We used the newly built LIVE Wild Compressed Video Quality Database to evaluate and compare the performance of the 1stepVQA model against other FR, RR and NR VQA models. 
It is first worth noting that while Difference Mean Opinion Score (DMOS) labels are typically supplied with existing quality research databases, those have available pristine undistorted videos as references. 

However, as discussed in \cite{yu2019predicting}, given imperfect references, DMOS is unable to capture the absolute quality of videos, e.g. in the instances of UGC videos with pre-existing distortions as reference videos. 
Hence we use the original MOS as the subjective standards, since MOS represents the absolute perceived quality of each. 

We evaluated the relationships between predicted quality scores and MOS using SROCC, Pearson’s (linear) correlation coefficient (LCC) and the Root Mean Squared Error (RMSE). 
To compute LCC and RMSE, the predicted scores are passed through a logistic non-linearity before computing performance. 
SROCC measures the ranked correlation of the given samples, and does not require any remapping. 
Larger values of SROCC and LCC indicate better performance, while larger values of RMSE imply worse performance. 

The best way to evaluate the generality of training based algorithms is to conduct cross-database training and testing. 
However, in our situation, there is no similar database that is publicly available. 
Hence we were only able to test on the new database. 
To do this, we randomly divided the database into non-overlapping 80\% training and 20\% test sets, with no overlap of original content between sets. 
The same randomized splits were repeated over 1000 iterations to avoid biased results, and we report the median values of the results. 
When testing the models that do not need training, we only report the median values of each such model's results on the 1000 test sets, as with the training based models. 

For comparison, we tested against prominent FR and RR models: PSNR, MS-SSIM, FSIM, ST-MAD, VSI, and our previously developed 2stepQA, and NR models: NIQE, BRISQUE, V-BLIINDS, and TLVQM. 
We did not include any deep learning models because of the lack of enough training samples, although this remains an open and interesting research direction. 

Since we found that the 70th feature of TLVQM was the same on all of the videos, we excluded it when training the model.

\subsection{Comparisons Against Mainstream VQA Methods}
We first computed and compared the performance of 1stepVQA against several reference and NR VQA models, with the results shown in Table \ref{performance_all_model}. 
To determine whether there exists significant differences between the performances of the compared models, we conducted a statistical significance test. 
We used the distributions of the obtained SROCC scores computed over the 1000 random train-test iterations. 
The non-parametric Wilcoxon Rank Sum Test \cite{wilcoxon1945individual}, which compares the rank of two lists of samples, was used to conduct hypothesis testing. 
The null hypothesis was that the median for the row model was equal to the median of the column model at the 95\% significance level. 
The alternate hypothesis was that the median of the row was different from the median of the column. 
A value of `1' in the table represents that the row algorithm was statically superior to the column algorithm, while a value of `-1' means the counter result. 
A value of `0' indicates that the row and column algorithms were statistically indistinguishable (or equivalent). 
The statistical significance results comparing the performances of the compared VQA algorithms using SROCC and LCC are tabulated in Tables \ref{wilcoxon_inter_SROCC} and \ref{wilcoxon_inter_LCC} respectively. 
The significance tests show that 1stepVQA significantly outperformed most of the other models. 
One exception was FSIM, which did nearly as well using an expensive phase measurement model. 
It may be useful to consider similar phase coherency features tooled towards better efficiency.

\begin{table*}
\caption{Performances of The 1stepVQA Model against Various FR, RR and NR VQA Models on the LIVE Wild Compressed Video Quality Database. Italics Indicate NR Algorithms.}
\label{performance_all_model}
\resizebox{\textwidth}{!}{%
\begin{tabular}{|c|c|c|c|c|c|c|c|c|c|c|c|}
\hline
      & PSNR    & MS-SSIM & FSIM   & ST-MAD & VSI    & 2stepQA & \textit{NIQE} & \textit{BRISQUE} & \textit{V-BLIINDS} & \textit{TLVQM} & 1stepVQA \\ \hline
SROCC & 0.5084  & 0.7856  & 0.8778 & 0.8197 & 0.7813 & 0.8493  & 0.7150        & 0.7877           & 0.8276             & 0.8381         & 0.8839   \\ \hline
LCC   & 0.5074  & 0.7744  & 0.8776 & 0.8141 & 0.7806 & 0.8455  & 0.7149        & 0.7887           & 0.8228             & 0.8303         & 0.8809   \\ \hline
RMSE  & 11.2782 & 8.3797  & 6.3419 & 7.7239 & 8.1939 & 7.1600  & 9.1012        & 8.0702           & 7.8482             & 7.5766         & 6.3032   \\ \hline
\end{tabular}%
}
\end{table*}

\begin{table}
\caption{Results of One-Sided Wilcoxon Rank Sum Test Performed Between SROCC Values of The VQA Algorithms Compared In Table \ref{performance_all_model}. A Value Of "1" Indicates That The Row Algorithm Was Statistically Superior to The Column Algorithm; " $-$ 1" Indicates That the Row Was Worse Than the Column; A Value Of "0" Indicates That the Two Algorithms Were Statistically Indistinguishable. Italics Indicate NR Algorithms.}
\label{wilcoxon_inter_SROCC}
\resizebox{\columnwidth}{!}{%
\begin{tabular}{|c|c|c|c|c|c|c|c|c|c|c|c|}
\hline
                           & PSNR & MS-SSIM & FSIM & ST-MAD & VSI & 2stepQA   & \textit{NIQE} & \textit{BRISQUE} & \textit{V-BLIINDS} & \textit{TLVQM} & 1stepVQA \\ \hline
PSNR                       & 0    & -1      & -1   & -1     & -1  & -1        & -1            & -1               & -1        & -1             & -1       \\ \hline
MS-SSIM                    & 1    & 0       & -1   & -1     & 0   & -1        & 1             & 0                & -1        & -1             & -1       \\ \hline
FSIM                       & 1    & 1       & 0    & 1      & 1   & 1         & 1             & 1                & 1         & 1              & -1       \\ \hline
ST-MAD                     & 1    & 1       & -1   & 0      & 1   & -1        & 1             & 1                & 0         & -1             & -1       \\ \hline
VSI                        & 1    & 0       & -1   & -1     & 0   & -1        & 1             & 0                & -1        & -1             & -1       \\ \hline
2stepQA                    & 1    & 1       & -1   & 1      & 1   & 0         & 1             & 1                & 1         & 1              & -1       \\ \hline
NIQE                       & 1    & -1      & -1   & -1     & -1  & -1        & 0             & -1               & -1        & -1             & -1       \\ \hline
BRISQUE                    & 1    & 0       & -1   & -1     & 0   & -1        & 1             & 0                & -1        & -1             & -1       \\ \hline
V-BLIINDS                  & 1    & 1       & -1   & 0      & 1   & -1        & 1             & 1                & 0         & -1             & -1       \\ \hline
TLVQM                      & 1    & 1       & -1   & 1      & 1   & -1        & 1             & 1                & 1         & 0              & -1       \\ \hline
1stepVQA                   & 1    & 1       & 1    & 1      & 1   & 1         & 1             & 1                & 1         & 1              & 0        \\ \hline
\end{tabular}%
}
\end{table}

\begin{table}
\caption{Results of One-Sided Wilcoxon Rank Sum Test Performed Between LCC Values of The VQA Algorithms Compared In Table \ref{performance_all_model}. A Value Of "1" Indicates That The Row Algorithm Was Statistically Superior to The Column Algorithm; " $-$ 1" Indicates That the Row Was Worse Than the Column; A Value Of "0" Indicates That the Two Algorithms Were Statistically Indistinguishable. Italics Indicate NR Algorithms.}
\label{wilcoxon_inter_LCC}
\resizebox{\columnwidth}{!}{%
\begin{tabular}{|c|c|c|c|c|c|c|c|c|c|c|c|}
\hline
                           & PSNR & MS-SSIM & FSIM & ST-MAD & VSI & 2stepQA & \textit{NIQE} & \textit{BRISQUE} & \textit{V-BLIINDS} & \textit{TLVQM} & 1stepVQA \\ \hline
PSNR                       & 0    & -1      & -1   & -1     & -1  & -1      & -1            & -1               & -1        & -1             & -1       \\ \hline
MS-SSIM                    & 1    & 0       & -1   & -1     & -1  & -1      & 1             & -1               & -1        & -1             & -1       \\ \hline
FSIM                       & 1    & 1       & 0    & 1      & 1   & 1       & 1             & 1                & 1         & 1              & -1       \\ \hline
ST-MAD                     & 1    & 1       & -1   & 0      & 1   & -1      & 1             & 1                & 0         & -1             & -1       \\ \hline
VSI                        & 1    & 1       & -1   & -1     & 0   & -1      & 1             & -1               & -1        & -1             & -1       \\ \hline
2stepQA                    & 1    & 1       & -1   & 1      & 1   & 0       & 1             & 1                & 1         & 1              & -1       \\ \hline
NIQE                       & 1    & -1      & -1   & -1     & -1  & -1      & 0             & -1               & -1        & -1             & -1       \\ \hline
BRISQUE                    & 1    & 1       & -1   & -1     & 1   & -1      & 1             & 0                & -1        & -1             & -1       \\ \hline
V-BLIINDS                  & 1    & 1       & -1   & 0      & 1   & -1      & 1             & 1                & 0         & -1             & -1       \\ \hline
TLVQM                      & 1    & 1       & -1   & 1      & 1   & -1      & 1             & 1                & 1         & 0              & -1       \\ \hline
1stepVQA                   & 1    & 1       & 1    & 1      & 1   & 1       & 1             & 1                & 1         & 1              & 0        \\ \hline
\end{tabular}%
}
\end{table}

\subsection{Limits of Reference Models}
\label{VMAF_biase}
As mentioned earlier, reference models can only measure the perceptual fidelity of a compressed video relative to a possibly distorted reference video, but do not take into account the pre-existing quality of the reference. 

From Table \ref{SROCC_FSIM_MOS_level}, it may be generally observed that for reference VQA models, as the compression becomes heavier, SROCC also increases. 
The likely reason for this is that the quality of the compressed videos is dominated by the pre-existing distortions of the source video, on which the MOS varies more widely. 
When the compression is heavier, the qualities of the tested videos becomes dominated by compression artifacts, which are more predictable than the panoply of possible pre-existing distortions. 

It is interesting that the training based models follow an opposite performance trend against the amount of compression. 
They are better able to distinguish the quality of lightly compressed videos from each other (within each each range of light compression) more accurately. 
When the compression is heavy, very poor quality videos from each other becomes more difficult for the trained models. 

\begin{table}
\caption{SROCC between Several VQA Model Scores and MOS for Each Compression Level. Italics Indicate NR Algorithms.}
\label{SROCC_FSIM_MOS_level}
\centering
\begin{tabular}{|c|c|c|c|c|c|}
\hline
Compression level & 1       & 2       & 3       & 4       & All    \\ \hline
PSNR              & -0.2818 & -0.1818 & 0.0364  & 0.1000  & 0.5084 \\ \hline
MS-SSIM           & -0.0091 & 0.0818  & 0.2273  & 0.3636  & 0.7856 \\ \hline
FSIM              & -0.0545 & 0.2455  & 0.3636  & 0.5091  & 0.8778 \\ \hline
\textit{BRISQUE}  & 0.6727  & 0.5364  & 0.1909  & 0.1091  & 0.7877 \\ \hline
1stepVQA          & 0.5909  & 0.5636  & 0.3727  & 0.1364  & 0.8839 \\ \hline
\end{tabular}
\end{table}

\subsection{Feature Combination Performance Evaluation}
\label{1stepVQA_intra_section}
We also conducted a series of feature studies to evaluate other possible feature group combinations. 
First, we studied the role of multiscale, since it generally improves prediction performance \cite{mittal2012no} (videos and distortions are multiscale), but it comes with some additional expense. 
Hence we compared against the results of using NVS features computed only at the original scale. 
Second, we tested the efficacy of using displaced frame differences in other directions than the four diagonal orientations. 
Specifically, we tested performance using frame differences in the cardinal directions: vertical and horizontal instead of diagonal, along with the non-displaced frame difference. 
We also studied the efficacy of using all eight displacement directions. 
Overall, we analyzed three extended versions of 1stepVQA: 
\begin{itemize}
  \item 1stepVQA-I: NVS features only at original scale.  
  \item 1stepVQA-II: NVS features computed on displaced differences in cardinal directions only, along with non-displaced frame differences. 
  \item 1stepVQA-III: NVS features computed on displaced frame differences in all eight adjacent directions, as well as the non-displaced frame difference. 
\end{itemize}
We tested the three models and report the obtained results in Table \ref{1stepVQA_intra}. 
Table \ref{wilcoxon_1stepVQA_feat_extend} shows the obtained statistical significance test comparing the performance of 1stepVQA against the three extended versions of 1stepVQA. 
It is evident that including or modifying the feature groups does not enhance performance, and not using multiscale reduces it. 
Indeed, 1stepVQA was statistically superior to all the other feature combinations.

\begin{table}
\caption{Performances of The 1stepVQA Model And Three Extended 1stepVQA Model, with Different Feature Groups. }
\label{1stepVQA_intra}
\resizebox{\columnwidth}{!}{%
\begin{tabular}{|c|c|c|c|c|}
\hline
      & 1stepVQA & 1stepVQA-I & 1stepVQA-II & 1stepVQA-III \\ \hline
SROCC & 0.8791   & 0.8468     & 0.8584      & 0.8619       \\ \hline
LCC   & 0.8760   & 0.8472     & 0.8538      & 0.8575       \\ \hline
RMSE  & 6.4595   & 7.0470     & 6.9029      & 6.8583       \\ \hline
\end{tabular}%
}
\end{table}

\begin{table}
\caption{Results of One-Sided Wilcoxon Rank Sum Test Performed Between SROCC Values of The VQA Algorithms Compared In Table \ref{1stepVQA_intra}. A Value Of "1" Indicates That The Row Algorithm Was Statistically Superior to The Column Algorithm; " $-$ 1" Indicates That the Row Was Worse Than the Column; A Value Of "0" Indicates That the Two Algorithms Were Statistically Indistinguishable.}
\label{wilcoxon_1stepVQA_feat_extend}
\resizebox{\columnwidth}{!}{%
\begin{tabular}{|c|c|c|c|c|}
\hline
             & 1stepVQA & 1stepVQA-I & 1stepVQA-II   & 1stepVQA-III \\ \hline
1stepVQA     & 0        & 1          & 1             & 1            \\ \hline
1stepVQA-I   & -1       & 0          & -1            & -1           \\ \hline
1stepVQA-II  & -1       & 1          & 0             & 0            \\ \hline
1stepVQA-III & -1       & 1          & 0             & 0            \\ \hline
\end{tabular}%
}
\end{table}

\subsection{Computational Complexity}

As compared with other VQA models, 1stepVQA is relatively fast and computes fewer features than most, and hence is very efficient. 
We also compared the overall computational complexity of 1stepVQA with the FR VQA model ST-MAD and the NR VQA models V-BLIINDS and TLVQM. 
In Table \ref{time_complexity}, we list the time required (in sec.) to compute each quality measure on a video from the LIVE Wild Compressed Video Quality Database. 
In Table \ref{number_features}, we list the number of features used by each of the training-based VQA models. 
As Tables \ref{time_complexity} and \ref{number_features} show, 1stepVQA is quite efficient. 

\begin{table*}
\caption{Comparison of Compute Times of Various VQA Measures on A Video from the LIVE Wild Compressed Video Quality Database on a Xeon E5 1620 v3 3.5GHz PC With 64 GB of RAM. Italics Indicate NR Algorithms.}
\label{time_complexity}
\resizebox{\textwidth}{!}{%
\begin{tabular}{|c|c|c|c|c|c|c|c|c|c|c|c|}
\hline
              & PSNR & MS-SSIM & FSIM & ST-MAD & VSI & 2stepVQA &  \textit{NIQE} &  \textit{BRISQUE} &  \textit{V-BLIINDS} &  \textit{TLVQM} & 1stepVQA \\ \hline
Time (sec.) & 22   & 64      & 95   &   3096 & 203 & 356      & 292  & 72      & 2746      & 246   & 95       \\ \hline
\end{tabular}
}
\end{table*}

\begin{table}
\caption{A Comparison of the Number of Features of Various Feature-Based VQA Models. Italics Indicate NR Algorithms.}
\label{number_features}
\resizebox{\columnwidth}{!}{%
\begin{tabular}{|c|c|c|c|c|c|}
\hline
                   & \textit{NIQE} & \textit{BRISQUE} & \textit{V-BLIINDS} & \textit{TLVQM} & 1stepVQA \\ \hline
Number of Features &  36           &  36              & 46                 & 75             & 36       \\ \hline
\end{tabular}
}
\end{table}

\section{Conclusion}
\label{conclusion}
We have presented a new VQA model, called 1stepVQA, that is designed to tackle the problem of assessing the quality of compressed videos given reference videos afflicted by pre-existing, authentic distortions. 
1stepVQA is computed using information from both the reference and compressed videos, but does not require pristine reference videos. 
It uses features that capture the deviations of spatial and spatial-temporal statistical regularities caused by the presence of pre-existing in-capture distortions as well as post-capture compression artifacts. 
Our method utilizes both NFS features, as well as NVS features extracted from displaced frame differences. 
We find that 1stepVQA is able to outperform mainstream VQA models, while requiring relatively fewer features, making it more efficient and easy to implement. 
We also developed a significant new database for this problem, called the LIVE Wild Compressed Video Quality Database, which contains both UGC reference videos and many compressed versions of them, along with subjective quality labels assigned to them as the outcome of a human study.


%





\ifCLASSOPTIONcaptionsoff
  \newpage
\fi



%
%
%

\bibliographystyle{IEEEtran}
\bibliography{bare_adv}{}

%

\end{document}